\documentclass[twocolumn, aps, prx, longbibliography, groupedaddress, superscriptaddress]{revtex4-2}

\usepackage{xcolor}
\usepackage{amsmath}
\usepackage{amssymb}
\usepackage{graphicx}
\usepackage[english]{babel}\addto\captionsenglish{}
\usepackage{textcomp}
\usepackage{braket}
\usepackage{hyperref}
\usepackage{float}
\usepackage{mathtools}
\usepackage{dsfont}
\usepackage{lineno}
\usepackage{comment}
\usepackage{diagbox} % for diagonal split cell

\newcommand{\fref}[1]{Fig.~\ref{#1}}
\newcommand{\Fref}[1]{Figure~\ref{#1}}

\newcommand{\dgr}{^{\dagger}}

\newcommand{\enum}[1]{\mathit{e}^{#1}}

\begin{document}
\title{Raman anyon-Hubbard model in a 1D Bose lattice gas}

\author{Josep Cabedo}
\affiliation{Departament de F\'isica, Universitat Aut\`{o}noma de Barcelona, E-08193 Bellaterra, Spain}
\author{Claudio Iacovelli}
\email[Contact author: ] {claudio.iacovelli@icfo.eu}
\affiliation{Departament de F\'isica, Universitat Aut\`{o}noma de Barcelona, E-08193 Bellaterra, Spain}
\affiliation{ICFO - Institut de Ciencies Fotoniques, The Barcelona Institute of Science and Technology,
Av. Carl Friedrich Gauss 3, 08860 Castelldefels (Barcelona), Spain}

\author{Leticia Tarruell}
\affiliation{ICFO - Institut de Ciencies Fotoniques, The Barcelona Institute of Science and Technology,
Av. Carl Friedrich Gauss 3, 08860 Castelldefels (Barcelona), Spain}
\affiliation{ICREA, Pg. Llu\'is Companys 23, 08010 Barcelona, Spain}
\author{Alessio Celi}
\email[Corresponding author: ]{alessio.celi@uab.cat}
\affiliation{Departament de F\'isica, Universitat Aut\`{o}noma de Barcelona, E-08193 Bellaterra, Spain}

\begin{abstract}
The anyon-Hubbard model is the paradigmatic lattice model of one-dimensional (1D) anyons, in which chiral interactions make bosons equivalent to particles with fractional statistics. Its continuum limit, the chiral BF theory, emerges as the low-energy description of a Raman-coupled Bose gas with imbalanced interactions. Here, we show that loading such a gas into a 1D optical lattice realizes the lattice counterpart of this theory, which we name the Raman anyon-Hubbard model. The Raman-dressed bosons acquire density-dependent Peierls phases when tunneling through occupied sites. For a small statistical phase, the model reduces to the anyon-Hubbard model, while beyond this limit it defines a distinct lattice anyon model with unbounded rather than periodic statistical interactions. Numerically, we show that the dressed gas displays the defining properties of 1D anyons, namely a density-dependent asymmetry of the ground-state momentum distribution, interaction-induced asymmetric transport and chiral bound states. 
Remarkably, these features are readily accessible with Raman-coupled $^{39}$K atoms, opening up the experimental exploration of 1D anyons beyond the few-particle regime.
\end{abstract}

\maketitle

\section{Introduction}
One-dimensional anyons are currently at the focus of both theoretical and experimental research. Like the more studied two-dimensional (2D) anyons, they escape the dichotomy between bosons and fermions by obeying generalized commutation relations \cite{Leinaas_NCB_1977_fractional, Goldin_1981_JMP_fractional, Wilczek_PRL_fractional_1982, greiter_arXiv_2022_fractional}, while promising to be more easily accessible in experiments. 
A variety of microscopic models has been introduced to describe 1D anyons \cite{Haldane_PRL_1991, Ha_PRL_1994_1D_fractional, Murthy_PRL_1994_1D_frac, Kundu_PRL_1999, Batchelor_PRL_2006_1D_anyons, Calabrese_PRB_2007_1D_anyons, delCampo_PRA_2008_1D_anyons, Hao_pra_2008, Harshman_2020_anyons_three_body_AoP, Harshman_2022_top_exch_stats_1D, Nagies_2023_beyond_braid_arXiv, Hidalgo_pra_2025_two_1D_anyons}, which also provide effective descriptions of strongly correlated systems, from quantum spin chains \cite{Haldane_PRL_1991} to the edges of fractional quantum Hall liquids \cite{Wen_prb_1990_chiral_LL}, and have been explored as a resource for quantum computation \cite{Tosta_pra_2019_1D_fermionic_anyons_QC}. In one dimension, particles cannot circumvent each other, and their exchange necessarily proceeds through a collision. Anyonic statistics is therefore intimately tied to interactions, bringing these models within reach of the quantum simulation platforms where interactions can be finely tuned. In this work, we propose a feasible experimental scheme to access the many-body physics of a 1D anyon model with ultracold atoms.

The best known and studied lattice model for 1D anyons is the anyon-Hubbard model (AHM), where the anyonic behavior stems from density-dependent tunneling, with a phase set by the statistical phase $\theta$ that interpolates between bosons ($\theta=0$) and pseudofermions ($\theta=\pi$). The AHM has been predicted to exhibit a rich variety of exotic properties. Its ground state features an asymmetric momentum distribution \cite{Hao_pra_2009, Hao_PRA_2012_1D_hc_anyons, Tang_NJP_2015_GS_1DAHM} and Friedel oscillations \cite{Strater_PRL_2016_Friedel_anyons}, while the statistical phase drives transitions between distinct quantum phases \cite{Keilmann_NatComs_2011_1DAHM_optical_lattice, Greschner_prl_2015_1D_AHM, Arcila_PRA_2016_crit_points_AHM}. In the presence of nearest-neighbor interactions, tuning the statistics from bosons to fermions yields a rich phase diagram, where gapped and superfluid phases, the latter including pair superfluids, meet at a multicritical line \cite{Bonkhoff_prl_2025_extended_AHM}. Out of equilibrium, the model exhibits an asymmetric density expansion \cite{Wang_PRA_2014_anyons_q_walks, Liu_PRL_2018_anyon_walks}. At the few-body level, chiral bound states form even in the absence of onsite interactions \cite{Zhang_PRA_2017_anyon_bound_states}, and the symmetries and integrability of the model have been recently classified, revealing exactly solvable doublon states hidden in the continuum of scattering states \cite{Bonkhoff_arXiv_2026_symmetry_integrability}.

Despite its substantial theoretical interest, the experimental realization of the model with ultracold atoms still poses a great challenge. Recently, the asymmetric momentum distribution of hardcore 1D anyons was observed in a strongly interacting quantum gas through an impurity travelling in a spin-charge-separated Tonks-Girardeau gas \cite{Dhar_Nature_2025_anyonization}. A realization of the AHM for two particles in an optical lattice was reported before \cite{kwan2023realization}, leveraging a Floquet scheme based on multicolor lattice-depth modulation \cite{Cardarelli_PRA_2016_multicolor_depth_modulation} to engineer the required density-dependent tunnelings \cite{Keilmann_NatComs_2011_1DAHM_optical_lattice}. Building on this scheme, two-body ground states of the model have recently been prepared adiabatically, revealing the pseudo-fermionization of 1D anyons and the chiral nature of their bound states \cite{Bakkali_arXiv_2026_chiral_binding}. While such an approach gives access to arbitrary anyonic phases, the mapping onto the AHM holds within a restricted Fock space. Moreover, coherence times are strongly limited by the heating mechanisms associated to the Floquet driving of many-body systems \cite{weitenberg_natphys_2021_floquet}.

In fact, the footprints of 1D anyons, namely the asymmetric expansion of the system and the formation of chiral solitons, were first observed in the continuum, in a Raman-coupled Bose gas with imbalanced interactions \cite{Froelian_2022}. This system realizes, at the quantum level \cite{Chisholm_prr_2022_CBF}, the chiral BF theory \cite{Rabello_PLB_1995_gauge_1D_anyons, Rabello_prl_1996_1D_anyons, Aglietti_prl_1996, Jackiw_JNMP_1997, Griguolo1998}. This topological gauge theory was originally proposed to define 1D anyons through a dimensional reduction of Chern-Simons theory \cite{ZhangKivelson1989, JackiwPi, Ezawa2008}, the topological gauge theory describing anyons in two dimensions. Indeed, the chiral BF theory is the field theory corresponding to the Kundu model \cite{Kundu_PRL_1999}, and is the continuum limit of the 1D AHM \cite{Bonkhoff_PhysRevLett_2021_linear_anyons}.

Here we reverse the latter statement and show that the lattice counterpart of the chiral BF theory is realized by a Raman-coupled gas loaded into an optical lattice, providing a natural experimental platform to study strongly interacting 1D anyons. As in the continuum \cite{Chisholm_prr_2022_CBF}, the interplay of Raman coupling and imbalanced interactions leads to chiral interactions for the Raman-dressed bosons, i.e., atoms in momentum-dependent superpositions of the two coupled internal states, in the form of current-density terms, or equivalently of correlated tunnelings with a density-dependent phase. These statistical interactions define an effective anyon model on the lattice that we term the Raman anyon-Hubbard model. For small density-dependent tunneling phases, it reduces to the standard 1D AHM, while beyond this limit it defines a distinct model. Indeed, since the density-dependent correction enters the tunneling amplitude linearly rather than as a phase factor, the strength of its statistical interactions is unbounded, instead of periodic as in the AHM. The Raman anyon-Hubbard model and the AHM share key defining features of a 1D anyon model: an asymmetric momentum distribution in the ground state, the asymmetric particle transport and the formation of chiral bound states, which stem directly from effective chiral interactions of Raman-dressed bosons as in the chiral BF theory \cite{Froelian_2022}. Beyond the AHM limit, however, the gap of the chiral bound states grows monotonically with the strength of the statistical interactions, a feature with no counterpart in the AHM. Remarkably, since the Raman anyon-Hubbard model emerges from a low-energy effective description of a Raman-coupled Bose gas, its realization is neither hindered by many-body heating processes nor restricted to a truncated Fock space, unlike the Floquet realizations of the AHM. Our Raman approach thus complements the Floquet realizations, which give access to arbitrary statistical phases, by opening the experimental exploration of the many-body regime of 1D anyons with ultracold atoms.
\begin{figure*}[t!]
\centering
\includegraphics[width=0.8\linewidth]{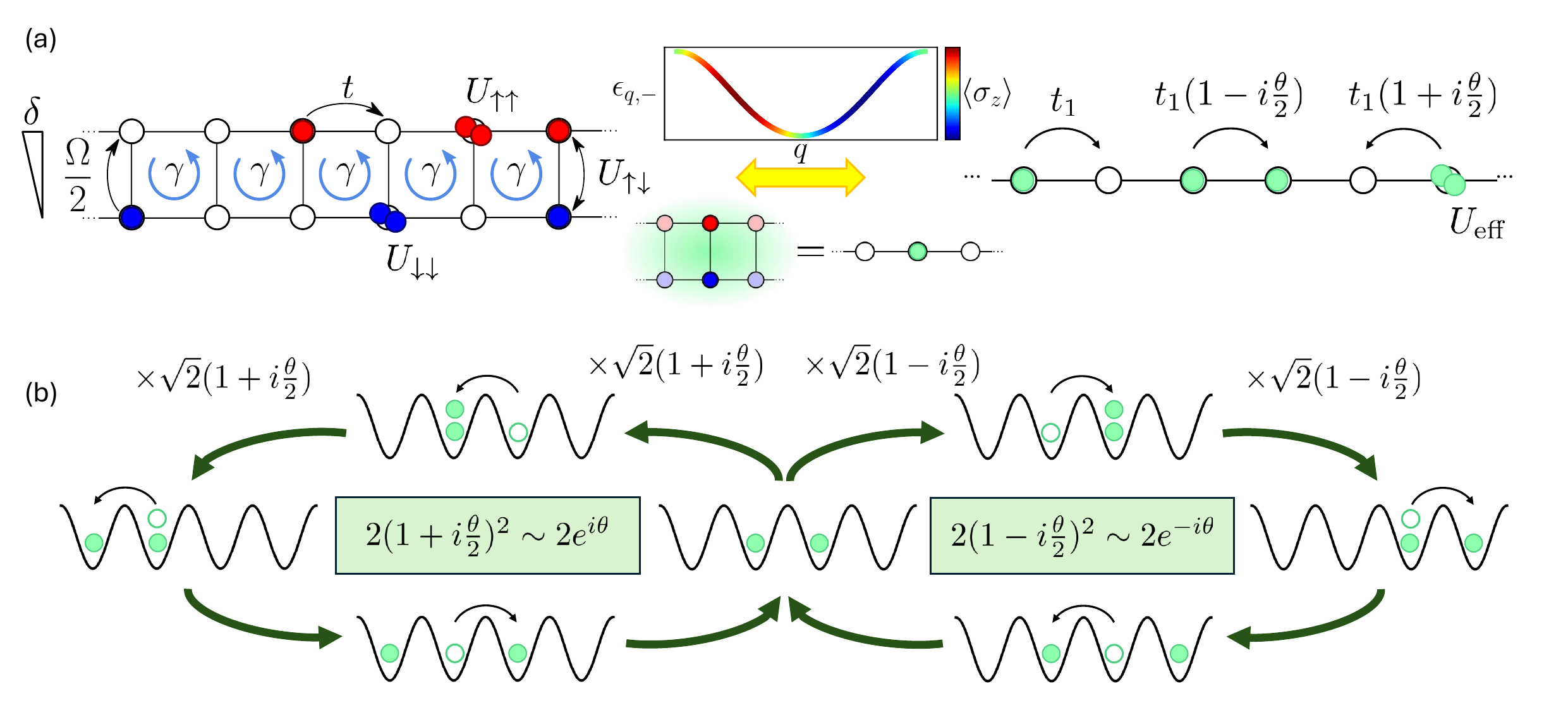}
\caption{(a) Left: a Raman-coupled spin-$1/2$ gas in an optical lattice realizes a bosonic flux ladder \eqref{eq_ham_flux_ladder}, with intra- and inter-leg tunnelings $t$ and $\Omega/2$, inter-leg offset $\delta$, intra-leg (inter-leg) interactions $U_{\uparrow\uparrow}$ and $U_{\downarrow\downarrow}$ ($U_{\uparrow\downarrow}$), and magnetic flux $\gamma$. Right: truncated lower-band Hamiltonian \eqref{eq_lower_band_Ham_theta}, describing spinless dressed bosons with onsite interactions $U_\mathrm{eff}$ and density-dependent complex tunnelings. Upper inset: energy $\epsilon_{q,-}$ of the lower single-particle band of the flux ladder as a function of quasimomentum $q$ for strong $\Omega$, with the color scale showing its spin polarization $\langle\sigma_z\rangle$, i.e., the spin texture that gives rise to effective chiral interactions (see main text). Lower inset: schematic representation of the lower-band Wannier modes (dressed bosons), consisting of a superposition of the upper and lower leg spin states with nonvanishing amplitudes in the adjacent sites. (b) Interferometric picture of the exchange of two adjacent dressed bosons \cite{kwan2023realization}. Displacing the pair by one site through a doubly occupied site (upper paths) or without overlap (lower paths) closes a loop in Fock space. For imbalanced interactions ($U_{\uparrow\uparrow} \neq U_{\downarrow\downarrow}$), tunneling onto or off an occupied site carries a factor $\sqrt{2}(1\pm i\theta/2)$, with $\sqrt{2}$ the bosonic enhancement and the sign set by the direction of motion. The two paths thus differ by $2(1\pm i\theta/2)^2$, whose phase reproduces the AHM exchange phase $\pm\theta$ only to first order in $\theta$.}
\label{Fig_lower_band_map}
\end{figure*}

The work is structured as follows. We introduce our treatment of the Raman-coupled 1D Bose lattice gas in Sec.~\ref{sec_chiral}. We show that, for strong Raman couplings, the interplay between the Raman coupling and the imbalanced interactions of the coupled internal states of the atoms gives rise to effective chiral interactions, parametrized by the coupling $\theta$, for Raman-dressed bosons. The resulting Raman anyon-Hubbard model displays a statistical phase proportional to $\theta$. For small $\theta$, the statistical phase is exactly $\theta$ and the Raman anyon-Hubbard model coincides with the AHM. Beyond this limit, it defines a new lattice anyon model with a density-dependent statistical phase. In Sec.~\ref{sec_props}, we numerically explore the properties of the Raman anyon-Hubbard model. We show that the ground state exhibits a density-dependent shift in the momentum distribution that scales with $\theta$. We also show that quantum walks of Raman-dressed bosons can reveal the effective statistical phase via spatial asymmetric transport of interferometric origin. To conclude the Section, we compute the two-particle spectrum, and confirm the presence of chiral bound states of Raman-dressed bosons. Such gapped states appear in the spectrum even in the absence of effective contact interactions. Next, in Sec.~\ref{sec_exp}, we discuss the experimental realization of the effective model with a Raman-coupled $^{39}$K lattice system. Finally, we briefly recap and draw our conclusions and outlook in Sec.~\ref{sec_conc}.
\section{Raman-coupled lattice: effective chiral interactions}\label{sec_chiral}

We consider an ultracold Bose gas loaded in a 1D optical lattice, deep enough to ensure the validity of the tight-binding approximation. We consider the gas to be further subjected to a pair of Raman beams that nearly-resonantly couple two internal states. These two states act as an effective spin-$1/2$ degree of freedom of the gas. As originally proposed in \cite{Celi-2014}, the different spin states can be conveniently reinterpreted as lattice sites along a synthetic dimension \cite{Boada-12}. In this picture, the Raman-coupled gas is effectively described as a spinless boson model on a ladder geometry, where each spin corresponds to a leg of the ladder, and the plaquettes are pierced by a static magnetic flux $\gamma$ (see \fref{Fig_lower_band_map}(a) left). The system is governed by the flux ladder Hamiltonian 
\begin{align}\label{eq_ham_flux_ladder}
H = &\sum_{j,\sigma}\left(-t e^{-i\gamma \sigma} a_{j+1,\sigma}^\dagger +\frac{\Omega}{4} a_{j,-\sigma}^\dagger - \frac{\delta}2 \sigma a_{j,\sigma}^\dagger \right)a_{j,\sigma} + \text{H.c.} \cr
& + \sum_{j,\sigma} n_{j,\sigma} \left( \frac{U_{\sigma,\sigma}}{2}(n_{j,\sigma}-1)+\frac{U_{\sigma,-\sigma}}{2} n_{j,-\sigma} \right).
\end{align}
Here, $a_{j\sigma}^\dagger (a_{j\sigma})$ creates (annihilates) a boson in the $j_\mathrm{th}$ site of the lattice with spin $\sigma=\pm1/2$. The coefficient $t$ is the tunneling rate along the lattice, $\Omega$ is the Raman Rabi frequency which controls the onsite coupling of the two spins, that is, the rung coupling of the ladder, and $\delta$ is the two-photon Raman detuning. Finally, the atoms experience onsite two-body interactions of strength $U_{\sigma,\sigma^{\prime}}$, which depend on the different spin-spin scattering lengths. Here, $n_{j,\sigma}=a_{j,\sigma}^\dagger a_{j,\sigma}$ is the density in site $j,\sigma$. Throughout the text, we will refer to Hamiltonian \eqref{eq_ham_flux_ladder} as the flux ladder model. We will use $\sigma$ to denote both the spin states $\sigma=\uparrow,
\downarrow$ and the corresponding spin quantum numbers $\sigma = \pm 1/2$. Throughout this Section, we set $\delta=0$.

To unveil the effective anyonic physics, we consider imbalanced interactions, $U_{\uparrow\uparrow} \neq U_{\downarrow\downarrow}$, and restrict ourselves to the strong Raman coupling regime $\Omega \gg t$, where the two single-particle dispersion bands are well separated by an energy gap $\sim \Omega$. Assuming a dilute gas where the mean interaction energy per particle is smaller than $\Omega$, the many-body problem can be captured by projecting the flux ladder onto the lower dispersion band \cite{Cabedo_epjd_2020,Barbiero_Cabedo_PRR_2023_frustrated}. Further, we consider a moderate flux $\gamma$ that gives a single-minimum dispersion as in \fref{Fig_lower_band_map}(a). Then, under these assumptions, the effective lower-band Hamiltonian displays chiral interactions, i.e., a current-density term:
\begin{align}\label{eq_lower_band_Ham_low_dens_not2}
H_{-}\! =\! \!\sum_{j}\!\Big[ &(-t_1 b_{j+1}^\dagger b_{j}\! +\! \mathrm{H.c.})\! + \frac{U_\mathrm{eff}}2\! :\!n_j n_j\!: \!+ t_1\theta  :\bar{j}_j n_j\!:\! \Big].
\end{align} 
Here, $b_j^\dagger (b_j)$ are the bosonic operators for the lower-band Wannier modes, which describe the Raman-dressed bosons on the lattice and which we simply call dressed particles. These modes are tightly localized along the spatial dimension, while they are delocalized along the synthetic one, being composed of a superposition of $\uparrow$ and $\downarrow$ spin states (see inset in \fref{Fig_lower_band_map}(a)). Moreover, $::$ denotes normal ordering, $n_j= b_j^\dagger b_j$ is the  dressed-particle density and $\bar{j}_j = (j_j + j_{j-1})/2$ averages the corresponding current operator $j_j =  i( b_{j+1}\dgr b_j -  b_j\dgr b_{j+1})$ over adjacent sites. The effective tunneling is directly determined by the shape of the lower band, 
\begin{equation}
    t_1 \simeq t\cos(\gamma/2),
\end{equation}
up to corrections of order $(t/\Omega)^2$, while the contact and chiral interaction coefficients read, respectively, 
\begin{align}
&U_\mathrm{eff} \simeq (U_{\uparrow\uparrow} + U_{\downarrow\downarrow} + 2U_{\uparrow\downarrow})/4,  \label{eq_Ueff_pert}\\
&\theta \simeq \tan(\gamma/2)(U_{\downarrow\downarrow} - U_{\uparrow\uparrow})/(2\Omega),\label{eq_theta}
\end{align}
plus corrections coming from second-order processes. These leading terms, which stem from the spin texture of the lower dispersion band and from the imbalance of the interactions, have a form similar to the one encountered in bulk Raman-coupled gases \cite{Chisholm_prr_2022_CBF}. 
Indeed, in both cases the spin composition of the dressed particles changes with momentum (see the color scale in the upper inset
of \fref{Fig_lower_band_map}(a)), so that spin-dependent interactions lead to effective momentum-dependent collisions, that is, current-density interactions.
The corrections from second-order processes, involving collisions with virtual particles in the upper band, renormalize the couplings and ensure the quantitative agreement of the effective model with the full flux ladder of Eq.~\eqref{eq_ham_flux_ladder}, as verified in Sec.~\ref{sec_props}. As in the bulk, the Raman coupling also induces a second-order correction to the effective mass of the particle. In the lattice, it manifests with the appearance of a second-neighbor tunneling $t_2 \propto t^2/\Omega$. The full derivation of these analytical parameters and the complete many-body Hamiltonian are detailed in Appendix~\ref{Appendix_A_lower_band_model}.
  
\subsection{The Raman anyon-Hubbard model}

The anyonic properties that stem from the current-density contribution in Hamiltonian \eqref{eq_lower_band_Ham_low_dens_not2} may be better understood by regrouping the chiral interacting terms into the kinetic part of the Hamiltonian. By doing so, the Hamiltonian can be conveniently rewritten as
\begin{align}\label{eq_lower_band_Ham_theta}
H_{-}\! =\!\! \sum_{j}\!\bigg[ &(-t_1 b_{j+1}^\dagger (1\! - i\theta \bar{n}_j) b_{j}\! + \mathrm{H.c.})\! + \frac{U_\mathrm{eff}}2 n_j( n_j\!-\!1) \bigg].
\end{align}
where we have defined the averaged density $\bar{n}_j = (n_j+n_{j+1})/2$. 
We refer to Hamiltonian \eqref{eq_lower_band_Ham_theta} as the \textit{Raman anyon-Hubbard model}. Indeed, from equation \eqref{eq_lower_band_Ham_theta}, we see that a dressed particle acquires a complex phase when it tunnels through an occupied adjacent site, as depicted in \fref{Fig_lower_band_map}(b). The sign of the phase acquired depends on the moving direction, a feature of manifestly anyonic nature. 
Indeed, to first order in $\theta$, the tunneling term in the Hamiltonian~\eqref{eq_lower_band_Ham_theta} can be approximated as a density-dependent Peierls phase:
\begin{align}\label{eq_lower_band_AHM}
H_{-} \sim \sum_{j}\bigg[ &-(t_1 b_{j+1}^\dagger \enum{-i\theta{\bar{n}_j}} b_{j} + \mathrm{H.c.}) + \frac{U_\mathrm{eff}}2 n_j( n_j\!-\!1) \bigg],
\end{align}
which shares the same functional form as the standard Jordan-Wigner transformed anyon-Hubbard model. Under a modified Jordan-Wigner transformation $a_j = \enum{i (\theta/2) \left(\sum_{k<j}n_k - \sum_{k>j}n_k\right)} b_j$, the operators $(a_i, a_j^\dagger)$ satisfy the proper anyonic commutation relations $a_i a_j ^ \dagger - \enum{i \theta \mathrm{sgn}(i-j)} a_j ^ \dagger a_i  = \delta_{ij}$ and are governed by the AHM $H_{-} = -\sum_{j}  ( t_1
 a_{j+1}^\dagger a_{j}  + \mathrm{H.c.}) + (U_\mathrm{eff}/2)\sum_j n_j(n_j - 1)$.

For small $\theta$, the model thus reduces to the standard AHM \eqref{eq_lower_band_AHM}. \Fref{Fig_lower_band_map}(b) illustrates this correspondence for two adjacent particles, whose exchange in 1D proceeds through a doubly occupied site and closes a loop in Fock space \cite{kwan2023realization}: the phase accumulated along the loop reproduces the AHM exchange phase only to first order in $\theta$. Beyond this limit, Eq. \eqref{eq_lower_band_Ham_theta} defines a distinct lattice anyon model, where the statistical interactions grow without bound with $\theta$, in contrast to the AHM, where they are periodic.
As we show in Appendix~\ref{Appendix_B_statistics}, the anyonic correspondence nonetheless survives for arbitrary $\theta$, and describes anyons that tunnel with a density-dependent amplitude and whose statistical phase depends on the local density. For two particles which do not share a bond it takes the exact value $\arctan\theta$, growing monotonically with the chiral coupling and saturating at $\pi/2$. 

These analytical results rely on the lower-band projection, valid for $\Omega \gg t, U_{\sigma\sigma'}$. In the next Section, we test them against numerical simulations of the full flux ladder \eqref{eq_ham_flux_ladder}, and show that i) it is accurately described by the Raman anyon-Hubbard model \eqref{eq_lower_band_Ham_theta}; ii)
it displays the key defining features of 1D anyons, which we compare with those of the AHM \eqref{eq_lower_band_AHM} both at small $\theta$, where the two models coincide, and beyond.
\begin{figure}[t!]
\centering
\includegraphics[width=\linewidth]{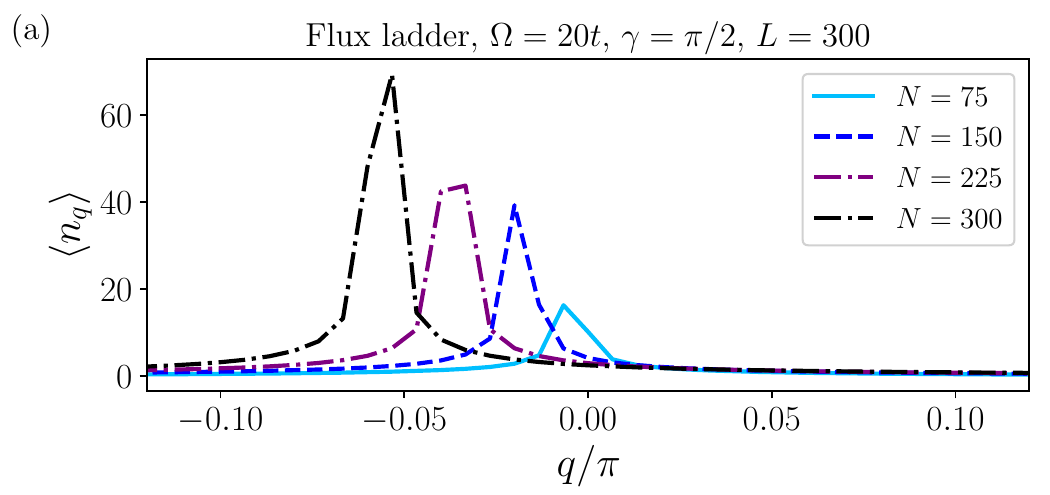}\\[1ex]
\includegraphics[width=\linewidth]{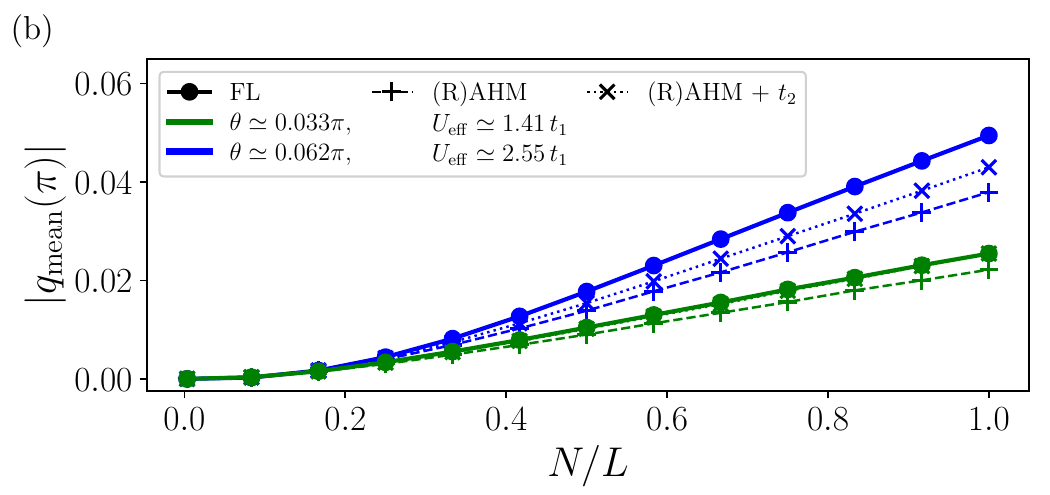}
\caption{Asymmetric momentum distribution. (a) Quasimomentum distribution of the ground state of the flux ladder model \eqref{eq_ham_flux_ladder} for different total numbers of particles $N$. The ladder is set to length $L=300$, with $\Omega = 20t$, $\gamma = \pi/2$ and $\delta = 0$. Interactions are set to $U_{\uparrow\uparrow} = 0.08t$, $U_{\downarrow\downarrow} = 15t$, and $U_{\uparrow\downarrow} = -0.37t$. (b) Mean value of the ground-state quasimomentum as a function of the particle filling $N/L$, for the interactions of (a) (blue) and for $U_{\uparrow\uparrow} = 0.025t$, $U_{\downarrow\downarrow} = 5t$, and $U_{\uparrow\downarrow} = -0.125t$ (green), with the effective parameters, including second-order corrections, given in the legend. Circles show the flux ladder (FL), $+$ markers the Raman anyon-Hubbard model \eqref{eq_lower_band_Ham_theta} and the AHM \eqref{eq_lower_band_AHM}, which coincide at this scale and are jointly labeled (R)AHM, and $\times$ markers the same models supplemented by the second-neighbor tunneling $t_2 \simeq -0.035\,t_1$, common to both sets. The shift grows with the filling, as expected from a density-dependent gauge field, and the effective models capture it at low filling. Adding $t_2$, i.e., the larger effective mass of the dressed bosons, extends the agreement to all fillings for the weaker interactions, while for the stronger ones higher-order corrections make it only qualitative at large $N/L$.}
\label{Fig_momentum_shifts}
\end{figure}

\section{Properties of the Raman anyon-Hubbard model}\label{sec_props}
The Raman anyon-Hubbard model \eqref{eq_lower_band_Ham_theta} derived in the previous Section holds for strong Rabi couplings and low atomic densities, and reduces to the AHM \eqref{eq_lower_band_AHM} for small effective statistical phases. In this Section, we challenge this effective description by numerically exploring the flux-ladder Hamiltonian \eqref{eq_ham_flux_ladder} and the effective model \eqref{eq_lower_band_Ham_theta} for large but experimentally feasible Rabi coupling and density up to one boson per rung, using the AHM as a benchmark for the anyonic character of the Raman-coupled gas. 
We find that the anyonic features are experimentally accessible in the flux ladder and that the Raman anyon-Hubbard model quantitatively explains them once second-order corrections are taken into account. Throughout the Section, the values of the different interaction parameters $U_{\sigma\sigma^\prime}$ of the flux ladder are chosen to match the s-wave scattering lengths of the ground state of $^{39}$K atoms at different accessible bias magnetic fields, as detailed in Sec. \ref{sec_exp} (see \fref{Fig_experimental_setup} and Table~\ref{tab_scattering_lengths} therein).

\subsection{Momentum asymmetry}\label{sec_momentum}

A remarkable feature of the AHM \eqref{eq_lower_band_AHM} is the asymmetric momentum distribution in its ground state \cite{Hao_pra_2009, Hao_PRA_2012_1D_hc_anyons, Tang_NJP_2015_GS_1DAHM}. The peak in the momentum distribution shifts away from the origin, with the size and the direction of the shift depending on the value of $\theta$. This behavior can be understood with a mean-field argument. The density-dependent gauge field shifts both the minimum of the single-particle dispersion and the peak of the momentum distribution by $\theta n$, where $n$ is the average density.   

We numerically test this property with density-matrix-renormalization-group (DMRG) \cite{White-prl-1992, Itensor, Itensor_r03} simulations of the flux ladder model \eqref{eq_ham_flux_ladder}, which we compare to the Raman anyon-Hubbard model (RAHM) of Eq.~\eqref{eq_lower_band_Ham_theta} and to its AHM limit \eqref{eq_lower_band_AHM}. In \fref{Fig_momentum_shifts}, we plot the quasimomentum distribution of the ground state in regimes where we expect the lower-band model to hold. Here, $q = kd$ is the adimensional quasimomentum, where $d$ is lattice spacing. Indeed, we observe the expected shifts in the distribution, with their strength depending on the value of the effective statistical phase $\theta$ and on the atomic density.

\begin{figure*}[t!]
\centering
\includegraphics[width=0.99\linewidth]{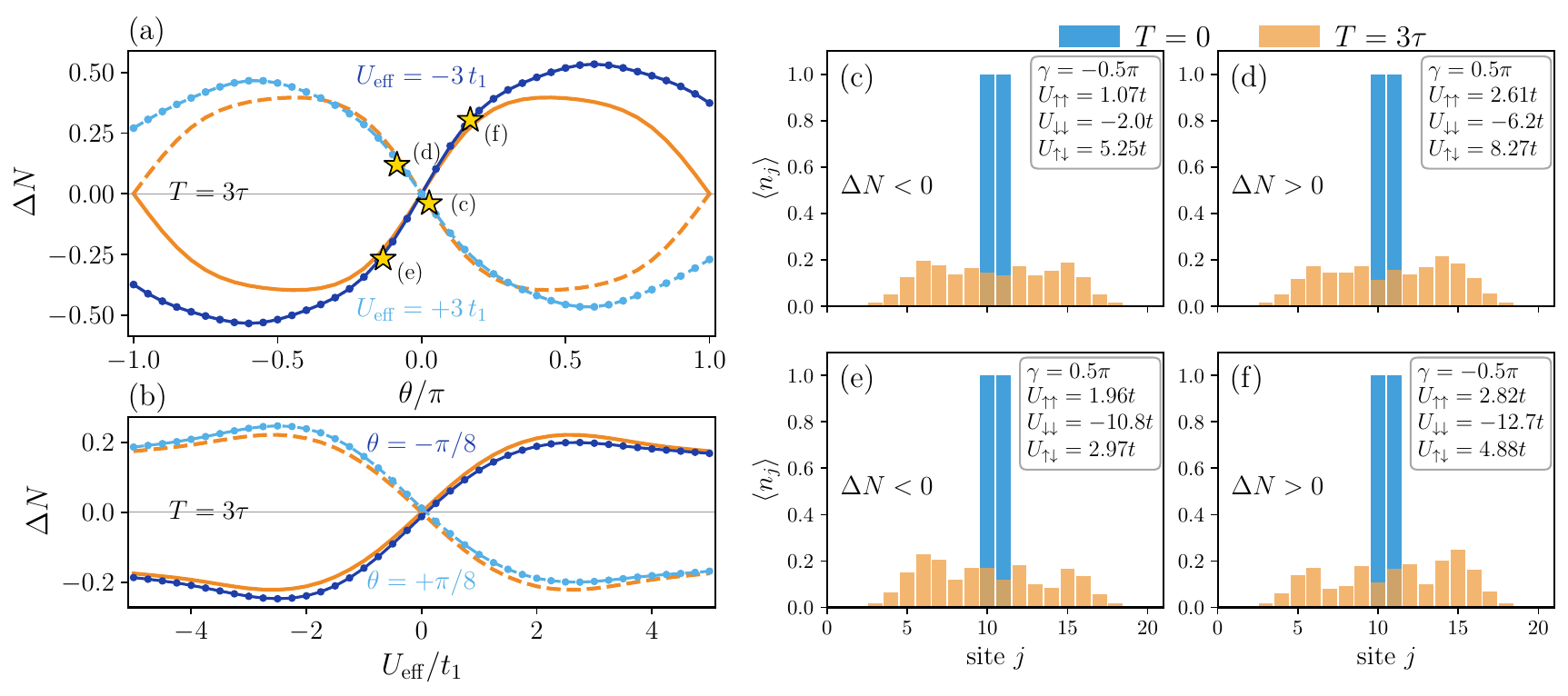}
\caption{Asymmetric transport. (a) Left-right population imbalance $\Delta N = \sum_{j=1}^{L/2} (n_{j+L/2}-n_{j})$ after free expansion of a two-particle state as a function of $\theta$. The particles are initialized at adjacent sites at the center of a $L=20$ chain, with $\ket{\psi(0)} = b_{10}^\dagger b_{11}^\dagger\ket{0}$, and evolve under the Raman anyon-Hubbard model \eqref{eq_lower_band_Ham_theta} supplemented by the second-neighbor tunneling $t_2$, for a time $T=3\tau$, where $\tau=\hbar/t_1$ is the tunneling time. We fix $U_\mathrm{eff} = 3 t_1$ (dashed light blue line) and $U_\mathrm{eff} = -3 t_1$ (solid dark blue line). In both cases, $t_2 = -0.035 t_1$. For comparison, orange lines show the results obtained with the AHM \eqref{eq_lower_band_AHM}. The yellow stars denote the values obtained with the full flux ladder model \eqref{eq_ham_flux_ladder}, which correspond to the simulations plotted in panels (c)-(f). (b) Left-right population imbalance after free expansion as a function of $U_\mathrm{eff}$, for a fixed value of $\theta = 0.125\pi$ (dashed light blue line) and $\theta = -0.125\pi$ (solid dark blue line). Orange lines show the results for the AHM. (c)-(f) Density distributions corresponding to the flux ladder results shown in (a). The densities are plotted at times $T=0$ and $T=3\tau$, and for different values of $U_{\sigma\sigma'}$ and $\gamma$. In all cases, $\Omega = 20t$ and $\delta = 0$. The expansion is asymmetric only for nonzero $\theta$ and $U_\mathrm{eff}$ (up to small $t_2$ corrections), with a direction set by their relative sign. It is most sensitive to $\theta$ around $\theta=0$, where all models coincide, and the flux ladder quantitatively follows the effective model.}
\label{Fig_asymmetric_expansions}
\end{figure*}

\Fref{Fig_momentum_shifts}(a) shows the quasimomentum profiles of the flux-ladder ground state for different particle fillings. The interaction coefficients $U_{\sigma\sigma'}$ are chosen to have both $\theta > 0$ and $U_\mathrm{eff} > 0$, according to Eqs.~\eqref{eq_theta} and \eqref{eq_Ueff_pert}, respectively. The latter is necessary to avoid the collapse of the ground state in the many-body regime. As expected, a positive $\theta$ results in a negative shift of the momentum distribution, which grows with the number of particles. This density dependence is shown in \fref{Fig_momentum_shifts}(b) for two sets of parameters, together with the results of the Raman anyon-Hubbard model \eqref{eq_lower_band_Ham_theta} and of the AHM \eqref{eq_lower_band_AHM}, which are indistinguishable at these small values of $\theta$. At larger fillings, the flux ladder departs quantitatively from the effective models, with even larger shifts. These are largely determined by the increased effective mass of the dressed bosons, which in the lattice appears as the second-neighbor tunneling $t_2$. Indeed, in the presence of $t_2$ the lower-band dispersion with a density-dependent gauge field reads
\begin{align}
&-2t_1 \cos q + 2|t_2| \cos 2q + 2 t_1 \theta n \sin q \nonumber\\
&\qquad \simeq (t_1-4|t_2|)\Big(q+\frac{t_1\,\theta n}{t_1-4|t_2|}\Big)^2 + \mathrm{const.},
\end{align}
so that the mean-field shift $\theta n$ is enhanced by the factor $t_1/(t_1-4|t_2|) = \left[1-4\,(t/\Omega)\tan(\gamma/2)\sin(\gamma/2)\right]^{-1} \simeq 1.16$. Once supplemented by $t_2$, the effective models account for the shifts of the flux ladder for the weaker interactions (green), while for the stronger ones (blue) higher-order perturbative corrections to $\theta$ and $U_\mathrm{eff}$ start contributing as well (see Appendix~\ref{Appendix_A_lower_band_model}).

\subsection{Asymmetric transport}\label{sec_density_exp}

Another defining feature of 1D anyons is the asymmetric expansion under time evolution when onsite interactions are also present \cite{Wang_PRA_2014_anyons_q_walks, Liu_PRL_2018_anyon_walks}. This property was recently probed in the first experimental realization of the AHM for two particles \cite{kwan2023realization}. The asymmetry here has its origin in the explicit breaking of inversion symmetry in the Hamiltonian, and it is arguably more revealing of the anyonic nature of the particles. In the first place, while the momentum distribution is not invariant under Jordan-Wigner transformation from anyons to bosons, the position space density and the position space correlators are. Moreover, such an asymmetric particle transport directly distinguishes between the presence of static Peierls phases and the density-dependent ones that appear in Hamiltonian \eqref{eq_lower_band_AHM} \cite{Wang_PRA_2014_anyons_q_walks, Liu_PRL_2018_anyon_walks}. The distinction is clear, since, in the former scenario, one can always find a gauge transformation that restores inversion symmetry in the Hamiltonian. To explore this property in both the effective and the flux ladder model \eqref{eq_ham_flux_ladder}, we simulate quantum walks of two-particle states, evolved under the Raman anyon-Hubbard model \eqref{eq_lower_band_Ham_theta} supplemented by the second-neighbor tunneling $t_2$. The two particles are initially placed in adjacent sites at the center of a lattice of size $L=20$, that is, $\ket{\psi(0)} = b_{10}^\dagger b_{11}^\dagger\ket{0}$, which is inversion-symmetric. After time evolution, however, the states exhibit an asymmetric density profile, with the degree of asymmetry depending both on $\theta$ and on $U_\mathrm{eff}$. We quantify the asymmetry with the population imbalance between the right and the left halves of the lattice, $\Delta N = \sum_{j=1}^{L/2} (n_{j+L/2}-n_{j})$. We note that the profiles obtained after expansion are specific of the initial Fock state chosen. In fact, the asymmetry, and the role of both the statistical phase and the onsite interactions, can be understood through an interferometric interpretation of Fock state evolution (see \fref{Fig_lower_band_map}(b)), as detailed in \cite{kwan2023realization}. 

In \fref{Fig_asymmetric_expansions}(a) we plot $\Delta N$ as a function of $\theta$, after evolving the initial state $\ket{\psi(0)}$ under the Raman anyon-Hubbard model \eqref{eq_lower_band_Ham_theta} supplemented with second-neighbor tunneling $t_2$, for a total time $T=3\tau$, where $\tau=\hbar/t_1$ is the tunneling time. We show both the case for positive ($U_\mathrm{eff} = 3 t_1$) and negative ($U_\mathrm{eff} = -3 t_1$) effective onsite interactions, which lead, at a given $\theta$, to population imbalances of opposite sign. We set $t_2 = -0.035 t_1$ in order to match the results obtained with the full flux ladder model \eqref{eq_ham_flux_ladder}, indicated with yellow star-shaped dots. For comparison, we also show the results obtained with the AHM \eqref{eq_lower_band_AHM}. The two models progressively depart as $|\theta|$ grows: the imbalance of the AHM, periodic and odd in $\theta$, vanishes again at $\theta=\pm\pi$, whereas that of the Raman anyon-Hubbard model remains sizable, as its statistical interactions are unbounded.
We notice that the region with the highest sensitivity $\vert \partial \Delta N/\partial \theta\vert$ is found in the vicinity of $\theta = 0$, where the three models coincide. This property makes the probing of the asymmetric expansions accessible even at small values of $\theta$. Similarly, in \fref{Fig_asymmetric_expansions}(b) we show $\Delta N$ as a function of $U_\mathrm{eff}$, keeping $\theta$ constant. A nonzero $t_2$ leads to small shifts in the population imbalance, with $\Delta N$ taking a nonzero value at $U_\mathrm{eff}=0$. A finite $t_2$ makes the model equivalent to a triangular ladder \cite{Barbiero_Cabedo_PRR_2023_frustrated}, whose plaquettes are pierced by a gauge-invariant flux. This flux, however, is barely resolved by the dynamics, as the time needed to encircle a plaquette scales as $1/t_2$ and diverges as $t_2\rightarrow 0$, so that its effect remains small in the regimes considered here. The anyonic physics we probe is therefore left unaffected by the presence of a small second-neighbor tunneling. Finally, in \fref{Fig_asymmetric_expansions}(c)-(f) we plot the density profiles before and after the free expansions computed with the flux ladder model that correspond to the star-shaped points shown in \fref{Fig_asymmetric_expansions}(a). Each panel uses ladder parameters that match the values of $U_\mathrm{eff}$ and $\theta$ in (a), and shows a remarkable agreement between the full flux ladder and the effective model, also for large $\vert U_{\sigma \sigma'}\vert$.

\begin{figure*}[t!]
\centering
\includegraphics[width=0.99\linewidth]{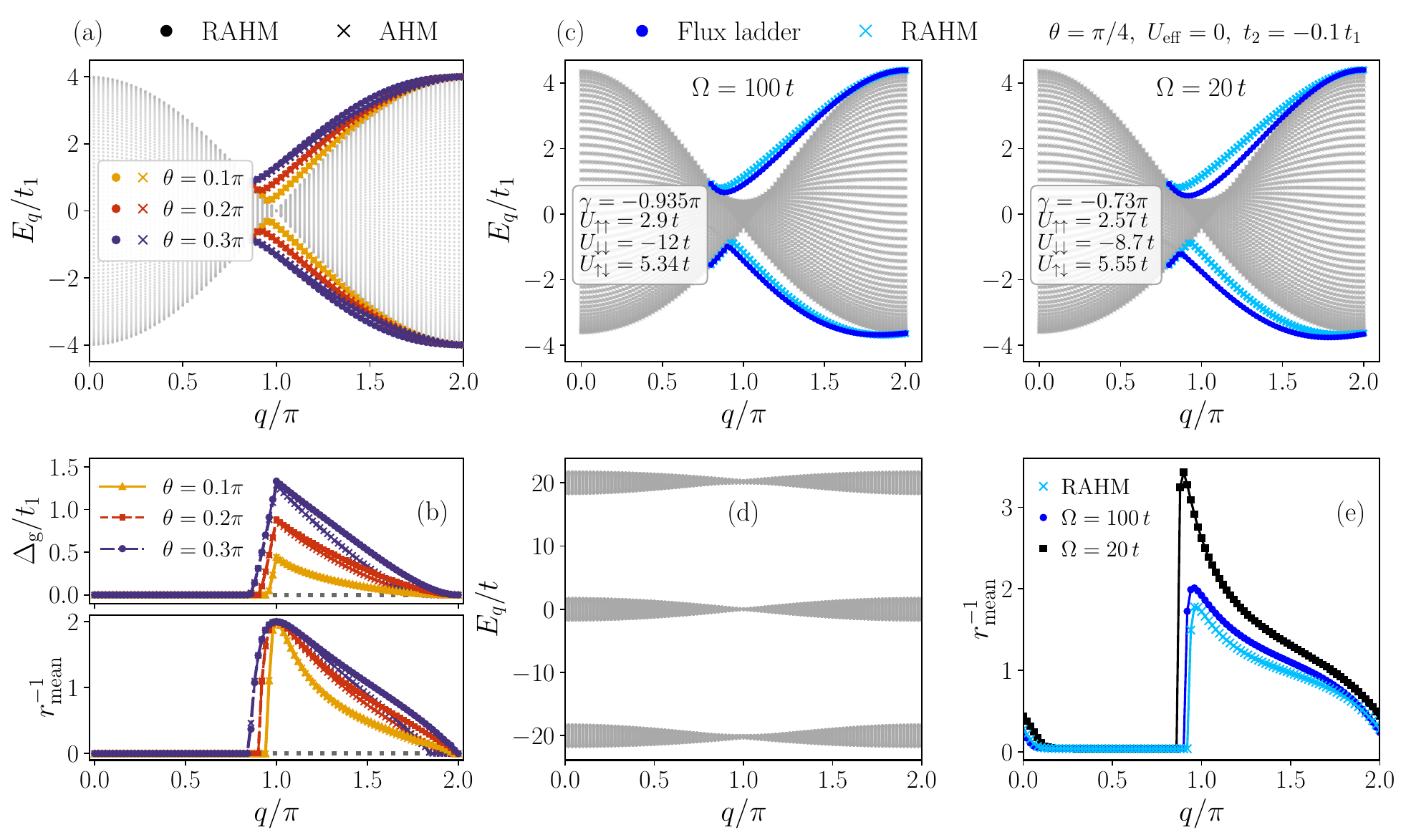}
\caption{Chiral bound states of the Raman anyon-Hubbard model. (a) Two-particle spectrum $E_q$ of the RAHM \eqref{eq_lower_band_Ham_theta} at $U_\mathrm{eff} = 0$ as a function of the center-of-mass momentum $q$, obtained by truncating the interparticle distance to $r_\mathrm{max}=100$. Colored dots show the upper and lower bound-state branches at $\theta = 0.1\pi$ (orange), $0.2\pi$ (red) and $0.3\pi$ (purple), and crosses of the same color the corresponding results for the AHM \eqref{eq_lower_band_AHM}. (b) Binding energy $\Delta_\mathrm{g}$ (upper panel) and inverse mean interparticle distance $r_\mathrm{mean}^{-1}$ (lower panel) of the lowest two-particle eigenstate as a function of $q$, extrapolated to $r_\mathrm{max} \rightarrow \infty$. Both quantities vanish for $\theta=0$ (gray dotted line). (c) Two-particle spectrum of the flux ladder model \eqref{eq_ham_flux_ladder} for $\Omega = 100 t$ (left) and $\Omega = 20 t$ (right), obtained with $r_\mathrm{max}=50$. Blue dots mark its bound states, which detach from the scattering continuum (gray), and light crosses show the RAHM supplemented by the second-neighbor tunneling $t_2$ of Appendix~\ref{Appendix_A_lower_band_model}. In both cases, $\delta = 0$, and $\gamma$ and $U_{\sigma\sigma'}$ are adjusted to give $\theta = 0.25\pi$, $U_\mathrm{eff} = 0$ and $t_2 = -0.1 t_1$, so that the RAHM results, which depend only on $\theta$ and $t_2/t_1$, are the same in both panels. Energies are scaled to $t_1$, and only the lower-band sector of the spectrum is shown. (d) Full two-particle spectrum of the noninteracting flux ladder at $\Omega = 20t$ and $\gamma = 0.73\pi$. (e) $r_\mathrm{mean}^{-1}$ of the lowest two-particle eigenstate for the flux ladder and the RAHM with the parameters of (c). At $U_\mathrm{eff}=0$, the statistical interactions alone bind the particles into two branches, below and above the scattering continuum, which exist only for $q\gtrsim\pi$: the lower branch has only negative group velocities $\mathrm{d}E_q/\mathrm{d}q$, the upper one only positive ones (a). Their gap grows with $\theta$ and exceeds that of the AHM at $\theta=0.3\pi$ (b). These chiral bound states persist in the flux ladder even at $\Omega = 20t$ (c,e), where the interaction energies approach the band gap.}
\label{Fig_bound_states}
\end{figure*}

\subsection{Chiral bound states}\label{sec_bound_states}

One-dimensional anyonic statistics can give rise to the formation of bound states of particle pairs \cite{Zhang_PRA_2017_anyon_bound_states, kwan2023realization,Bakkali_arXiv_2026_chiral_binding}. Notably, these gapped states may occur even in the absence of density-density interactions. They arise instead from the statistical interactions alone, which are encoded in the density-dependent tunneling of the Raman anyon-Hubbard model \eqref{eq_lower_band_Ham_theta}. Next, we explore its two-particle spectrum and show that it indeed hosts two branches of chiral bound states. In contrast to the AHM \eqref{eq_lower_band_AHM}, whose spectrum is periodic in $\theta$, we find that the binding energy of these states increases monotonically with $\theta$, which makes the bound-state spectrum a direct probe of the regime beyond the AHM. Finally, we show that such bound states and their properties are preserved back in the full flux ladder model \eqref{eq_ham_flux_ladder} for a wide range of parameters.

In order to treat the two-body problem, we introduce the two-particle Fock basis $\ket{j_1,\sigma_1,j_2,\sigma_2}$, where $j_1$ ($j_2)$ labels the position of the leftmost (rightmost) particle. The indices $\sigma_{1,2}$ naturally extend the basis to spin space for the case of the flux ladder model. With this notation, we introduce a family of bases labeled by the center of mass momentum $q$
\begin{equation}
\ket{e_{q, r}^{\sigma,\sigma^\prime}}=\frac{1}{\sqrt{L}}\sum_{j}e^{iq\left(j+\frac{r}{2}\right)}\ket{j,\sigma, j+r,\sigma^\prime}\label{eq:com_momentum_basis},
\end{equation}
where $r$ is the interparticle distance and $j+r/2$ the center of mass coordinate. By truncating the interparticle distance $r$ to a maximum value $r_\mathrm{max}$, the two-particle Hamiltonian can be numerically diagonalized (see Appendix~\ref{Appendix_C_TwoBodyH}). In \fref{Fig_bound_states}(a), we plot the two-particle eigenstates for both the Raman anyon-Hubbard model \eqref{eq_lower_band_Ham_theta} and the AHM \eqref{eq_lower_band_AHM} as a function of $q$. In both cases, we set $U_{\mathrm{eff}}=0$ and $r_\mathrm{max} = 100$. In grey, we plot the eigenvalues that correspond to the scattering solutions. In the thermodynamic limit, they are expected to form a continuum in the region $\vert E_q\vert\leq 4t_1\vert\cos(q/2)\vert$. The colored branches of eigenvalues correspond to gapped bound state solutions that depart from the continuum of scattering states. For small values of $\theta$, we observe a very good agreement between the two models, with larger quantitative differences as $\theta$ is increased and the Raman anyon-Hubbard model departs from its AHM limit. 

This difference is already visible at $\theta = 0.3\pi$, where the bound-state branches of the Raman anyon-Hubbard model lie further from the scattering continuum than those of the AHM. Indeed, since the strength of the chiral interaction is unbounded, the gap keeps increasing with $\theta$, whereas in the AHM it is periodic. Remarkably, the particles form bound pairs even in the absence of density-density interactions. The nature of these solutions is illustrated in \fref{Fig_bound_states}(b), where we plot the energy gap $\Delta_g$ and the mean inverse interparticle distance $r_\mathrm{mean}^{-1}$ in the lower energy eigenstate as a function of $q$, extrapolated to $r_\mathrm{max}\rightarrow \infty$. Around $q=\pi$, where the binding energy is the largest, the two particles are tightly confined to each other, with a mean distance of $\sim 0.5$ sites. Both the binding energy and $r_\mathrm{mean}^{-1}$ rapidly decay to zero as $q$ is decreased from $\pi$, as the ground state shifts from negative to positive group velocities $v_g(q) = \mathrm{d}E_q/\mathrm{d}q$. In contrast, the bound character of the states remains where $v_g < 0$. Such an asymmetry illustrates the markedly chiral nature of the bound states, a hallmark of the model that is closely related to the soliton solutions of the chiral BF theory in the continuum observed in \cite{Froelian_2022}. Indeed, neither the scattering states nor conventional density-density attractively- or repulsively-bound states show an asymmetry between left- and right-movers in their spectrum. We note that for negative values of $\theta$ the spectrum of both models is mirrored with respect to $q=\pi$, reverting the sign of the group velocities of the bound states. In the presence of repulsive (attractive) onsite interactions, the bound solutions are shifted upwards (downwards) in energy, and the spectrum loses the mirror symmetry between upper and lower branches about $E_q = 0$. Through this picture, one can explain the asymmetric transport described in the previous Section in terms of the bound-state solutions \cite{kwan2023realization}. For $U_\mathrm{eff}\neq 0$, the initial state projects asymmetrically to the lower and upper branches of bound states, and takes a nonzero averaged group velocity when chiral interactions are also present. 

Finally, we investigate the extent to which the chiral bound solutions predicted by the lower-band model persist back in the flux ladder Hamiltonian. The agreement between the two models is illustrated in \fref{Fig_bound_states}(c), where we compare the two-particle spectrum of the flux ladder model \eqref{eq_ham_flux_ladder} to that of the Raman anyon-Hubbard model \eqref{eq_lower_band_Ham_theta}, here supplemented by the second-neighbor tunneling $t_2$. We show two different sets of ladder parameters, all chosen to achieve the same effective couplings $\theta \simeq \pi/4$, $U_{\mathrm{eff}}\simeq 0$ and $t_2 \simeq  -0.1\ t_1$ in the lower-band model, according to Eqs.~\eqref{Seq_eff_tunnelings} and \eqref{Seq_Utheta_2ndpt} of Appendix~\ref{Appendix_A_lower_band_model}. In the left panel, we set $\Omega = 100t$, for which we observe a very good agreement between the two models. This agreement is consistent with the large gap $\sim \Omega$ between the single-particle bands, which is an order of magnitude greater than the interacting energies $U_{\sigma\sigma'}$, therefore rendering the lower-band theory very accurate. Still, a good qualitative agreement persists even for much smaller values of $\Omega$, where the interacting and non-interacting scales become closer in magnitude. Indeed, we can still distinguish two gapped branches that detach from the scattering continuum. The lower branch, in particular, appears to retain quantitatively similar eigenvalues. 

For a direct comparison between the two models, in \fref{Fig_bound_states}(c) we only show the lower-energy set of eigenvalues of the flux ladder model. This includes the scattering states where the two particles occupy single-particle states in the lower dispersion band, plus the two lower branches of bound states. In turn, all energies are shifted to center the lower-band scattering continuum around the origin, and scaled to the respective values of $t_1$. The full spectrum of the model in the absence of interatomic interactions is shown in \fref{Fig_bound_states}(d). We can clearly distinguish three two-particle scattering continua that are energetically well separated. From lower to upper, these correspond to states where two, one or zero particles occupy states in the lower single-particle band. Accordingly, they are separated by a gap that is close to the value of $\Omega$. Finally, in \fref{Fig_bound_states}(e) we show the inverse of the mean interparticle distance $r_\mathrm{mean}^{-1}$ in the lowest energy solutions plotted in \fref{Fig_bound_states}(c). Again, we observe an abrupt change in interparticle distance when the group velocity becomes negative, which shows that both the bound and the chiral nature of the gapped solutions persist in the flux ladder system. Importantly, these features are preserved even as the interaction energies $U_{\sigma\sigma'}$ approach the band gap. Being able to relax the lower-band condition and tune $\Omega$ to a large range of values can facilitate the detection of this property in a realistic experimental implementation. 
\section{Experimental considerations}\label{sec_exp}

\begin{figure}[t!]
\centering
\includegraphics[width=0.99\linewidth]{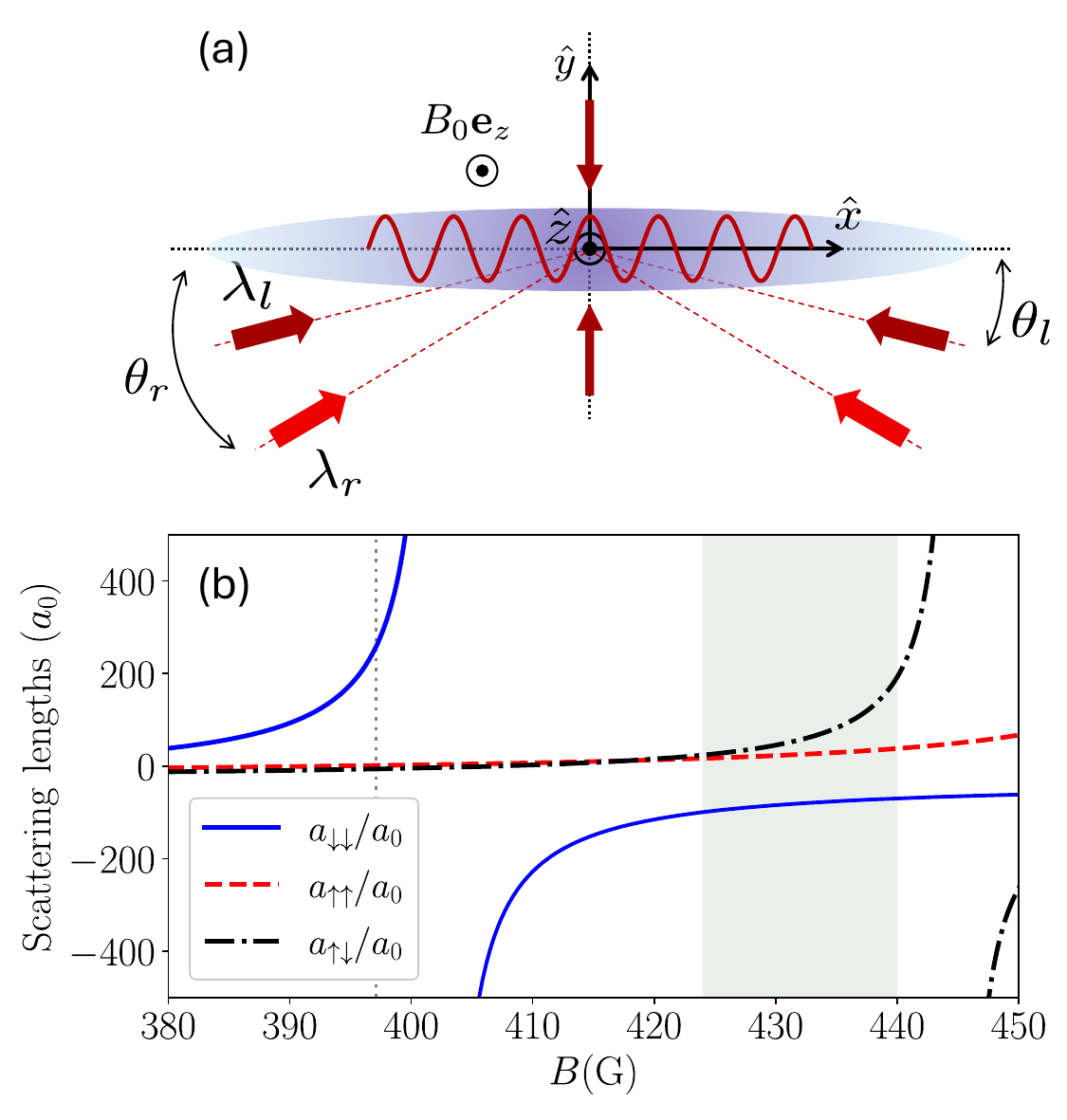}
\caption{(a) Experimental scheme. An ultracold gas is held in a 1D lattice, with tight confinement along the transversal directions. An additional pair of beams induce Raman transitions between two internal states of the atoms, which are energetically separated with a bias magnetic field $B_0$. The Raman momentum transfer is set along the lattice direction. (b) Scattering lengths $a_{\uparrow\uparrow}$, $a_{\uparrow\downarrow}$ and $a_{\downarrow\downarrow}$ as a function of the bias field strength $B_0$ for the levels of the ground-state manifold of $^{39}$K used as spin states, which at field $B_0=0$ correspond to $\ket{\uparrow} = \ket{F=1, m_F = 0}$ and $\ket{\downarrow} = \ket{F=1, m_F = 1}$. Here, $a_0$ is the Bohr radius. The vertical dashed line indicates the value considered for the results shown in \fref{Fig_momentum_shifts}, and the shadowed region includes the range of values used in Figs.~\ref{Fig_asymmetric_expansions} and \ref{Fig_bound_states}, which are listed in Table~\ref{tab_scattering_lengths}.}
\label{Fig_experimental_setup} 
\end{figure}

The flux ladder model \eqref{eq_ham_flux_ladder} can be realized with a Raman-coupled ultracold gas subjected to a deep 1D lattice potential. This approach facilitates achieving large rung couplings along the synthetic direction with low Raman intensities \cite{Stuhl-15, Mancini-15}. The lattice is characterized by a laser wavelength $\lambda_l$ and an opening angle $\theta_l$, which define the intersite lattice spacing $d = \lambda_l/(2\cos\theta_l)$. In turn, Raman beams of wavelength $\lambda_r$ are placed at an opening angle $\theta_r$, as depicted in \fref{Fig_experimental_setup}(a). The strongly-interacting regime of the lattice can be reached by setting a strong confinement along the transversal directions. Relevantly, the flux that pierces the plaquettes, given by  $\gamma = 2\pi \lambda_l\cos\theta_r/(\lambda_r\cos\theta_l)$, can be widely adjusted by tuning the values of both opening angles.

In order to achieve strong effective chiral interactions, we require an atomic species compatible with largely imbalanced interactions, and which is in turn suitable for Raman dressing. An ideal choice is $^{39}$K, especially the $F=1$ Zeeman sublevels, as shown in \cite{Froelian-thesis}. In particular, we set $\ket{\uparrow} \equiv \ket{F=1, m_F = 0}$ and $\ket{\downarrow} \equiv \ket{F=1, m_F = 1}$. The corresponding intra- and inter-state s-wave scattering lengths \cite{DErrico_2007_njpj_scattering_39K, Roy_prl_2013_scattering_39K} are shown in \fref{Fig_experimental_setup}(b). As illustrated, with this choice of atomic species and internal states, we can broadly tune the interaction parameters due to the presence of Feshbach resonances found at accessible bias magnetic fields of $\approx 400$\,G. The interaction parameters used in the previous Section are chosen in accordance, with the specific choices being listed in Table~\ref{tab_scattering_lengths}. At such values of the bias field, the atomic energy levels considered are found well within the Paschen-Back regime, and thus largely insensitive to magnetic field fluctuations, making the experimental realization robust. 

\begin{table}[t!]
    \centering
    \begin{tabular}{|c|c|c|c|c|}
        \hline
        & $B_0$\,(G) & $a_{\downarrow\downarrow}/a_0$  & $a_{\uparrow\downarrow}/a_0$ & $a_{\uparrow\uparrow}/a_0$ \\
        \hline
        Fig.\,2 & 397.1 & 257.31 & $-6.31$ & 1.31\\
        Fig.\,3c & 439.8 & $-70.05$ & 183.89 & 37.5\\
        Fig.\,3d & 436.3 & $-74.2$ & 98.95 & 31.28\\
        Fig.\,3e & 425 & $-96.41$ & 26.47 & 17.48\\
        Fig.\,3f & 427.5 & $-89.79$ & 34.5 & 19.93\\
        Fig.\,4c left & 428.6 & $-87.27$ & 38.8 & 21.09\\
        Fig.\,4c right & 431.3 & $-81.88$ & 52.26 & 24.22\\
        \hline
    \end{tabular}
    \caption{List of bias magnetic field values $B_0$ and corresponding scattering lengths for the $^{39}$K levels considered (see \fref{Fig_experimental_setup}), and that are compatible with the flux ladder parameters shown in
                Figs.~\ref{Fig_momentum_shifts}--\ref{Fig_bound_states}.}
    \label{tab_scattering_lengths}
\end{table}

A wide range of lattice and Raman dressing configurations is compatible with the regimes of the flux ladder model discussed in this work. Consider, for instance, a conventional red-detuned retro-reflected ($\theta_l = 0$) optical lattice of wavelength $\lambda_l=1064\;\mathrm{nm}$. The targeted values of the synthetic flux $\vert \gamma \vert \sim (n+1/2)\pi$ can then be realized with a pair of Raman beams at the tune-out wavelength $\lambda_r = 768.9$\,nm, placed with several compatible opening angles found within the range $0< \theta_r <\pi/2$ (in particular, at either $\theta_r \sim 0.14\pi$, $\sim 0.32\pi$ or $\sim 0.44\pi$). In these conditions, effective tunneling rates of $t_1/h \sim 100$\,Hz can be reached with lattice depths of $\sim 8 E_l$, with $E_l$ being the lattice single-photon recoil energy. Accordingly, the values of $U_{\sigma\sigma'}$ used in Sec.~\ref{sec_props} can be reached with additional transverse lattices with depths of a few tens of $E_l$. Finally, the lower-band physics explored in this work demands strong rung couplings, $\Omega \simeq 20t$ in most of Sec.~\ref{sec_props}. In the semi-synthetic realization of the flux ladder model, such regimes can be accessed with Raman intensities below the Raman recoil energy $E_r$, allowing for lifetimes of several tens of tunneling times \cite{Wei_2013, Froelian_2022}, well beyond the expansion time $T = 3\tau$ of Sec.~\ref{sec_density_exp}. 

\section{Conclusion}\label{sec_conc}    
In this work, we have shown that a 1D Raman-coupled bosonic lattice with imbalanced interatomic interactions realizes a lattice anyon model, which we call the Raman anyon-Hubbard model. This result establishes the lattice counterpart of the continuum anyonic physics that was experimentally realized in a Raman-coupled Bose gas \cite{Froelian_2022, Chisholm_prr_2022_CBF}, completing the bridge between the continuum and lattice descriptions of 1D anyons within a single experimental platform. Following a lower-band description of the many-body system, we found that the dressed bosons acquire density-dependent Peierls phases when tunneling through occupied sites, which are equivalent to current-density interactions. In the limit of small chiral couplings, the model reduces to the 1D anyon-Hubbard model, while beyond this limit it defines a genuinely distinct lattice anyon model with unbounded statistical interactions. Unlike the AHM, where the statistical phase is periodic in the chiral coupling, the statistical interactions in our model grow without bound. This regime has no counterpart in the standard AHM. It remains largely unexplored and could host qualitatively new many-body phases. Guided by this analytical map, we have numerically tested several key features of anyonic behavior in the original dressed system.

Taking the AHM as our benchmark, we have verified that the dressed gas displays three defining properties of 1D anyons. In the many-body regime, DMRG simulations reveal a shift in the momentum distribution of the ground state that scales with both the effective statistical phase $\theta$ and the particle filling. In free quantum walks of two particles, the effective chiral interactions give rise to an asymmetric density expansion, which highlights the explicit breaking of inversion symmetry and whose degree depends both on $\theta$ and on the effective onsite interactions. Finally, the two-particle spectrum hosts two branches of gapped chiral bound states, which arise purely from the statistical interactions and can be interpreted as the few-body limit of the chiral solitons that exist in the continuum theory and have been experimentally demonstrated in \cite{Froelian_2022}. Unlike in the AHM, their gap grows monotonically with $\theta$, which makes the two-body spectrum a direct probe of the unbounded statistical interactions. Remarkably, we find these properties to extend beyond the regimes where the analytical map is expected to be accurate: the chiral bound states persist even as the interaction energies $U_{\sigma\sigma'}$ approach the band gap, allowing the Raman coupling $\Omega$ to be tuned over a wide range of experimentally accessible values without destroying the anyonic character of the system. This robustness should facilitate the experimental detection of the anyonic properties of the gas across a broad range of experimental parameters. In particular, we propose a realistic implementation with $^{39}$K atoms, owing to the large tunability of the interatomic interactions in this species.

Our proposal sits at the crossroads of several active research directions. As a lattice realization of the chiral BF theory, it connects to the broader class of topological phenomena described by this gauge theory, including the chiral edge modes of fractional quantum Hall states \cite{Wen_prb_1990_chiral_LL}. In the continuum, the asymmetric expansion and chiral solitons that are characteristic of 1D anyonic physics were demonstrated in a Raman-coupled Bose gas \cite{Froelian_2022}, and anyonization of bosons was observed through the momentum distribution of an impurity traversing a Tonks-Girardeau gas \cite{Dhar_Nature_2025_anyonization}. On the lattice, Refs.~\cite{kwan2023realization, Bakkali_arXiv_2026_chiral_binding} realized the AHM for two particles using a Floquet scheme based on lattice-depth modulation, demonstrating pseudo-fermionization and the chiral nature of anyonic bound states. However, Floquet heating poses severe challenges in the many-body regime \cite{weitenberg_natphys_2021_floquet}, and the Fock space accessible to this approach is inherently restricted \cite{Cardarelli_PRA_2016_multicolor_depth_modulation}, which has so far confined these realizations to the two-particle sector. Our Raman scheme is free from both limitations: the effective model emerges from a static lower-band description, with no many-body heating and no restriction on the Fock space. It therefore complements the Floquet realizations, which access arbitrary statistical phases in the few-body regime, by opening the many-body regime of lattice anyons to experimental exploration.

Several directions are open for future work. A first is to map the ground-state phase diagram of the Raman anyon-Hubbard model at finite density. There, the chiral bound states identified in the two-body spectrum hint at the emergence of partially paired phases, because, as observed in \fref{Fig_bound_states}(c), the minimum of the energy dispersion of the two-body bound states is comparable to that of the scattering states. Partially paired phases are known to occur in the AHM at finite statistical angle \cite{Greschner_prl_2015_1D_AHM, Zhang_PRA_2017_anyon_bound_states}, and may thus persist at moderate couplings in our model as well; the enhanced stability of two-particle bound states induced by the second-neighbor tunneling corrections hints at a further enriched partially paired structure in the flux ladder. A second direction points towards two dimensions by coupling several one-dimensional wires with different chiral interactions, exploring the physics of density-dependent gauge fields in quasi-two-dimensional geometries such as density-driven Meissner-to-vortex phase transitions \cite{Greschner_ladder}. Our work provides, therefore, an alternative and robust pathway for probing 1D anyonic properties in state-of-the-art experiments with ultracold atoms. It paves the way for the experimental realization of these systems beyond the few-particle regime, opening new possibilities for the exploration of fractional statistics in low-dimensional quantum systems.

\begin{acknowledgments}
We thank C. A. Guerra Gutiérrez for the exchanges on the Quantum Algebra Mathematica package, which we used to verify the perturbative calculations.
\textit{Funding:} We acknowledge funding from the European Union (ERC CoG-101003295 SuperComp), the Spanish Ministry of Science, Innovation and Universities MCIU/AEI/10.13039/501100011033/FEDER, UE (projects MAPS PID2023-149988NB-C22 and Severo Ochoa CEX2024-001490-S at ICFO, and project MAPS PID2023-149988NB-C21 at UAB), the EU QuantERA project QUASIMODO, Generalitat de Catalunya (CERCA program), Fundació Cellex and Fundació Mir-Puig. C.I. acknowledges support from the Spanish Ministry of Science and Innovation MCIN/AEI/10.13039/501100011033 and ESF through the predoctoral contract PRE2021-099050. J.C. acknowledges support from the Spanish Ministry of Economic Affairs and Digital Transformation through the QUANTUM ENIA project call – Quantum Spain project.
\end{acknowledgments}
\section*{Data Availability}
The data that support the findings of this article will be made openly available. %are openly available~\cite{RAHM_data}.

%%%%%%%%%% Appendix %%%%%%%%%%
% \clearpage
\appendix
\onecolumngrid

\section{Derivation of the lower-band model}\label{Appendix_A_lower_band_model}

In this Appendix, we provide the explicit derivation leading from the full synthetic flux ladder Hamiltonian \eqref{eq_ham_flux_ladder} to the effective Raman anyon-Hubbard model \eqref{eq_lower_band_Ham_low_dens_not2}, with the systematic inclusion of second-order corrections in $1/\Omega$.

\subsection{Single-particle Hamiltonian and the dressed modes}\label{Ssec-lower_band_model}

The flux ladder model \eqref{eq_ham_flux_ladder} is translation-invariant. In the absence of inter-atomic interactions, the Hamiltonian is therefore block diagonal in orthogonal momentum subspaces, and we can write $H_{\text{n.i.}} = \sum_q H_q$. Here $q = kd$ is the adimensional quasimomentum, and $H_q$ reads
\begin{align}\label{Seq_square_hamq_long}
H_q = \sum_{\sigma} &\Bigg[ \left(4t\sin\left(\frac{\gamma}{2}\right)\sin(q) - \delta\right)\sigma \tilde{a}_{q,\sigma}\dgr \tilde{a}_{q,\sigma} + \frac{\Omega}{2}\tilde{a}_{q,-\sigma}\dgr \tilde{a}_{q,\sigma} - 2t\cos\left(\frac{\gamma}{2}\right)\cos(q) \tilde{a}_{q,\sigma}\dgr \tilde{a}_{q,\sigma}\Bigg],
\end{align}
where $\tilde{a}_{q,\sigma} = L^{-1/2} \sum_j \enum{-i q j} a_{j,\sigma}$ are the Fourier-transformed lattice modes. In matrix form, Hamiltonian \eqref{Seq_square_hamq_long} reads
\begin{equation}\label{SSeq_square_hamq}
H_q = \left(2t\sin\left(\frac{\gamma}{2}\right)\sin(q)  - \frac{\delta}{2}\right)\sigma_z  -2t\cos\left(\frac{\gamma}{2}\right)\cos(q) + \frac{\Omega}{2}\sigma_x,
\end{equation}
where $\sigma_j$ are the spin-1/2 Pauli matrices that act on the pseudospin space spanned by the pair of coupled internal states with quasimomentum $q$. By diagonalizing \eqref{SSeq_square_hamq}, we obtain the two single-particle dispersion bands $\epsilon_{q,\pm}$, given by 
\begin{equation}\label{Seq_bands}
\epsilon_{q,\pm} = -2t  \cos\left(\frac{\gamma}{2}\right)\cos(q) \pm \frac{\Omega}2\sqrt{1 + (\tilde{q} - \tilde{\delta})^2},
\end{equation}
where we conveniently define $\tilde{q} \equiv 4t\sin(\gamma/2)\sin(q)/\Omega$ and $\tilde{\delta} \equiv \delta/\Omega$,
and write the Hamiltonian in its diagonal form as
\begin{equation}\label{Seq_square_ham_ni}
H_\mathrm{n.i.} = \sum_{q,m=\pm} \epsilon_{q,m} \tilde{b}_{q,m}\dgr\tilde{b}_{q,m}.
\end{equation}
Here we introduce the eigenmodes $\tilde{b}_{q,\pm}\dgr$ of the non-interacting dressed Hamiltonian
\begin{equation}\label{Seq_bandmodes0}
\tilde{b}_{q,\pm}\dgr = \sum_{\sigma} \mathcal{U}_{\pm,\sigma}(q) \tilde{a}_{q,\sigma}\dgr
\end{equation}
 with $\mathcal{U}(q) = \enum{-i\sigma_y\theta_q/2}$
being the unitary transformation that relates the bare and the dressed basis (assumed to be real), where $
\cos(\theta_q) = (\tilde q -\tilde \delta)/\sqrt{1 + (\tilde q - \tilde\delta)^2}, \quad 0\le\theta_q \le\pi$. Explicitly,
\begin{equation}\label{Seq_rotation1}
\mathcal{U}_{+\uparrow}(q) = \mathcal{U}_{-\downarrow}(q) =  \sqrt{\frac{1 + \cos(\theta_q)}{2}}, \quad \mathcal{U}_{+\downarrow}(q) = - \mathcal{U}_{-\uparrow}(q) = \sqrt{\frac{1 -\cos(\theta_q)}{2}}.
\end{equation}
Finally, we introduce the Wannier mode operators of the lower and upper single-particle band basis, respectively, which read
\begin{equation}\label{Seq_ftbandmodes}
b_{j}\dgr = \frac{1}{\sqrt{L}}\sum_q\enum{-iqj}\tilde{b}_{q,-}\dgr = \sum_{j'} \left[ \left( \frac{1}{L} \sum_q e^{iq(j'-j)} \mathcal{U}_{-\uparrow}(q) \right) a_{j',\uparrow}\dgr + \left( \frac{1}{L} \sum_q e^{iq(j'-j)} \mathcal{U}_{-\downarrow}(q)\right) a_{j',\downarrow}\dgr \right].
\end{equation}
\begin{equation}\label{Seq_ftbandmodes_c}
c_{j}\dgr = \frac{1}{\sqrt{L}}\sum_q\enum{-iqj}\tilde{b}_{q,+}\dgr = \sum_{j'} \left[ \left( \frac{1}{L} \sum_q e^{iq(j'-j)} \mathcal{U}_{+\uparrow}(q) \right) a_{j',\uparrow}\dgr + \left( \frac{1}{L} \sum_q e^{iq(j'-j)} \mathcal{U}_{+\downarrow}(q)\right) a_{j',\downarrow}\dgr \right],
\end{equation}
and rewrite \eqref{Seq_square_ham_ni} as
\begin{equation}\label{Seq_square_ham_ni_dressed}
H_\mathrm{n.i.} = -\sum_j \sum_l \left( t_{l,-} b_{j+l}\dgr b_{j} + t_{l,+} c_{j+l}\dgr c_{j} \right),
\end{equation}
with $t_{l,\pm} = -L^{-1}\sum_q\enum{iql}\epsilon_{q,\pm}$ 
being the effective tunneling coefficients coupling the Wannier dressed states. 

In this work, we consider the regime where the rung tunneling $\Omega$ (the Raman coupling strength) is the dominant energy scale, and the two dispersion bands are well separated by a gap
\begin{equation}\label{Seq_bandgap}
\Delta\epsilon = \operatorname{min}(\epsilon_{q,+} - \epsilon_{q',-}) = \Omega\left[1 - 4\cos\left(\frac{\gamma}{2}\right)\frac{t}{\Omega} + O\left( \left( \frac{t+\delta}{\Omega}\right)^2  \right) \right].
\end{equation}
We can then safely neglect the highest-band dressed states and truncate Hamiltonian \eqref{Seq_square_ham_ni_dressed} to
\begin{equation}\label{Seq_square_ham_ni_truncated}
H_\mathrm{n.i.} \simeq -\sum_j \sum_l  t_{l} b_{j+l}\dgr b_{j},
\end{equation}
where we conveniently relabel $t_{l,-}$ by $t_{l}$. By taking the limit $L\rightarrow \infty$, and explicitly expanding the expression of $t_{l}$ in terms of $t/\Omega$, one finds
\begin{align}\label{Seq_tunnelingcoeff_truncated}
t_{l} = t\Bigg[&\left(\frac{\Omega}{2t}\sqrt{1+\tilde{\delta}^2} + \frac{2t\sin^2(\gamma/2)}{\Omega(1+\tilde{\delta}^2)^{3/2}} \right)\delta_{l,0} + \left(\cos(\gamma/2) - i\frac{\tilde{\delta}\sin(\gamma/2)}{\sqrt{1+\tilde{\delta}^2}}\right)\delta_{l,1} + \left(\cos(\gamma/2) + i\frac{\tilde{\delta}\sin(\gamma/2)}{\sqrt{1+\tilde{\delta}^2}} \right)\delta_{l,-1} \cr
&-\frac{t\sin^2(\gamma/2)}{\Omega(1+\tilde{\delta}^2)^{3/2}} \delta_{l,2} - \frac{t\sin^2(\gamma/2)}{\Omega(1+\tilde{\delta}^2)^{3/2}} \delta_{l,-2} \Bigg] + O\left( (t/\Omega)^2\right).
\end{align}
Thus, to first order in $t/\Omega$, the dressed Hamiltonian \eqref{Seq_square_ham_ni_truncated} simply reads
\begin{equation}\label{Seq_square_ham_ni_truncated_2}
H_\text{n.i.} \simeq \sum_{j}(-t_{1} b_{j+1}^\dagger b_{j} - t_2 b_{j+2}^\dagger b_{j} + \text{H.c.}),
\end{equation}
where we drop the constant onsite term $\propto t N$ and only nearest- and second-neighbor couplings are kept, with
\begin{equation}\label{Seq_eff_tunnelings}
\frac{t_{1}}{t} = \cos(\gamma/2) - i \frac{\tilde{\delta}\sin(\gamma/2)}{\sqrt{1+\tilde{\delta}^2}}, \qquad \frac{t_{2}}{t} = -\frac{t\sin^2(\gamma/2)}{\Omega(1+\tilde{\delta}^2)^{3/2}}.
\end{equation}
We note that in the limit $\Omega \gg t$, which is assumed in the main text, the second-neighbor tunneling vanishes $t_2\to 0$ and can be therefore neglected. In practice, however, $\Omega$ is experimentally bounded and $t_2$ remains small but finite, so we examine its effects in Section~\ref{sec_props}. This completes the derivation of the single-particle term in \eqref{eq_lower_band_Ham_low_dens_not2}.
\subsection{Interatomic interactions: effective current-density term}

In the bare basis, the interaction Hamiltonian reads
\begin{equation}\label{Seq_int_ham}
H_\mathrm{int} = \sum_{j, \sigma, \sigma^{\prime}}  \frac{U_{\sigma, \sigma^{\prime}}}2 : n_{j,\sigma} n_{j,\sigma^{\prime}} :,
\end{equation}
with $n_{j,\sigma}=a_{j,\sigma}^\dagger a_{j,\sigma}$ being the density in site $j,\sigma$. 
Since we consider the regime where the interacting energy per particle is smaller than the band gap, we will treat interactions within the lower-band truncation. As a starting point, we will expand the operators $n_{j,\sigma}$ in the dressed basis and truncate it to the lower band mode operators $b_{j}^\dagger$ and $b_{j}$. However, we will need to take into account virtual processes involving higher band states that have nonzero projection back in the lower band subspace, since they can add up to significant corrections to the lower-band model.

We can rewrite the bare basis modes in terms of the single-particle eigenmodes as
\begin{equation}\label{Seq_bare_modes_rotated}
a_{j,\sigma}\dgr = \sum_{l}\left( \lambda_{-,\sigma}^{(l)} b_{j+l}\dgr + \lambda_{+,\sigma}^{(l)} c_{j+l}\dgr\right),
\end{equation}
where we conveniently define the coefficients
\begin{equation}\label{Seq_lambda_coeffs_gen}
\lambda_{\pm,\sigma}^{(l)} = \frac{1}{L} \sum_q \enum{iql} \mathcal{U}^*_{\pm,\sigma}(q).
\end{equation}
By computing explicitly \eqref{Seq_lambda_coeffs_gen} using \eqref{Seq_rotation1}, we find the following nonzero coefficients $\lambda_{\pm,\sigma}^{(l)}$ to order $O(t/\Omega)$
\begin{align}\label{Seq_lambda_coeffs}
\lambda_{-\downarrow}^{(l)} & = \lambda_{+\uparrow}^{(l)} =  C_0 \delta_{l,0} - i \frac{\alpha}{2C_0}(\delta_{l,-1} - \delta_{l,1}) + O\left((t/\Omega)^2\right),\cr
\lambda_{-\uparrow}^{(l)} & = -\lambda_{+\downarrow}^{(l)} = -S_0 \delta_{l,0}  - i \frac{\alpha}{2S_0} (\delta_{l,-1} - \delta_{l,1}) + O\left((t/\Omega)^2\right),
\end{align}
and so
\begin{align}\label{Seq_bare_basis_truncated}
a_{j,\uparrow}\dgr &= C_0 c_{j}\dgr - S_0 b_{j}\dgr + i\frac{\alpha}{2 C_0} \left(c_{j+1}\dgr - c_{j-1}\dgr \right) + i\frac{\alpha}{2 S_0} \left(b_{j+1}\dgr - b_{j-1}\dgr \right) + O\left((t/\Omega)^2\right),\cr
a_{j,\downarrow}\dgr &= S_0 c_j\dgr + C_0 b_j\dgr - i\frac{\alpha}{2 S_0} \left(c_{j+1}\dgr - c_{j-1}\dgr \right) + i\frac{\alpha}{2 C_0} \left(b_{j+1}\dgr - b_{j-1}\dgr \right) + O\left((t/\Omega)^2\right).
\end{align}
Here, we introduce the coefficients $C_0 = \cos(\theta_0/2) = [(1 - \tilde\delta/\sqrt{1+\tilde{\delta}^2})/2]^{1/2}$, $S_0 = \sin(\theta_0/2) = [(1 + \tilde\delta/\sqrt{1+\tilde{\delta}^2})/2]^{1/2}$, and $\alpha = t\sin(\gamma/2)/[\Omega(1+\tilde\delta^2)^{3/2}]$.

We can now rewrite the interaction Hamiltonian \eqref{Seq_int_ham} in the dressed basis by simply substituting in equations \eqref{Seq_bare_basis_truncated}. As mentioned above, given the large gap that separates the dressed states of the two bands, we will treat the interactions that involve higher band modes as a perturbation $O\left(U_{\sigma,\sigma'}/\Omega\right)$ to the Hamiltonian truncated to the lower band modes. Let us start by computing the zero-order theory of the lower band interacting Hamiltonian by setting $\lambda_{+,\sigma}^{(l)} = 0$. With this truncation, and by defining the dressed-basis local density $n_j = b_j\dgr b_j$ and the following averaged current operator at sites $j$ and $j-1$
\begin{equation}\label{Seq_eff_current_j}
\bar{j}_j = \frac{j_j + j_{j-1}}2, \,\, \text{ with } \quad  j_j = i( b_{j+1}\dgr b_j -  b_j\dgr b_{j+1}),
\end{equation}
we can compactly write the onsite contributions from \eqref{Seq_int_ham} to the truncated lower-band interacting Hamiltonian as
\begin{align}\label{Seq_int_ham_contributions}
: n_{j,\uparrow}n_{j,\uparrow}: &= S_0^4 b_j\dgr b_j\dgr b_jb_j -2\alpha S_0^2 i\left[\frac{b_{j+1}\dgr b_j \dgr - b_{j-1} \dgr b_{j}\dgr }{2}b_jb_j - b_j\dgr b_j\dgr \frac{b_{j+1}b_{j} - b_{j-1}b_{j}}{2} \right] + O\left((t/\Omega)^2\right) \cr 
&= S_0^4 : n_j n_j : - 2\alpha S_0^2 : \bar{j}_j n_j: + O\left((t/\Omega)^2\right),\cr
: n_{j,\downarrow}n_{j,\downarrow}: &= C_0^4 b_j\dgr b_j\dgr b_jb_j +2\alpha C_0^2i\left[\frac{b_{j+1}\dgr b_j \dgr - b_{j-1} \dgr b_{j}\dgr }{2}b_jb_j - b_j\dgr b_j\dgr\frac{b_{j+1}b_{j} - b_{j-1}b_{j}}{2} \right] + O\left((t/\Omega)^2\right) \cr 
&= C_0^4 : n_j n_j : + 2\alpha C_0^2 : \bar{j}_j n_j: + O\left((t/\Omega)^2\right),\cr
: n_{j,\uparrow}n_{j,\downarrow}: &= S_0^2 C_0^2 b_j\dgr b_j \dgr b_j b_j + \alpha (S_0^2-C_0^2) i\left[\frac{b_{j+1}\dgr b_j \dgr - b_{j-1} \dgr b_{j}\dgr }{2}b_jb_j - b_j\dgr b_j\dgr \frac{b_{j+1}b_{j} - b_{j-1}b_{j}}{2} \right] + O\left((t/\Omega)^2\right)
\cr 
&= S_0^2 C_0^2 :n_j n_j: + \alpha (S_0^2-C_0^2) : \bar{j}_j n_j: + O\left((t/\Omega)^2\right).
\end{align}

By grouping terms, we retrieve the expression for the interaction Hamiltonian \eqref{Seq_int_ham} truncated to the lower band and to order $O(t/\Omega)$
\begin{equation}\label{Seq_int_ham_lower_band}
H_\mathrm{int}^{(0)} \simeq \sum_j \left( \frac{U_\mathrm{eff}^{(0)}}2 :n_j n_j : + \lambda^{(0)} : \bar{j}_j n_j : \right),
\end{equation}
where we have defined the coefficients
\begin{align}\label{Seq_int_ham_coeffs}
 U_\mathrm{eff}^{(0)} &= S_0^4 U_{\uparrow\uparrow} + C_0^4 U_{\downarrow\downarrow} + 2S_0^2 C_0^2 U_{\uparrow\downarrow},\cr
 \lambda^{(0)} &= \alpha\left( -S_0^2( U_{\uparrow\uparrow} -U_{\uparrow\downarrow}) + C_0^2( U_{\downarrow\downarrow} -U_{\uparrow\downarrow}) \right).
\end{align}
With this, we recover the functional form of the interacting part of the main-text Hamiltonian \eqref{eq_lower_band_Ham_low_dens_not2}, and in particular of the current-density term that encodes the chiral interactions, with the coefficients evaluated to leading order in $t/\Omega$.

\subsection{Second-order corrections from virtual upper-band collisions}
We now include back the collision processes that involve higher band states, treating them as a perturbation to Hamiltonian \eqref{Seq_int_ham_lower_band}, and use perturbation theory to obtain the effective Hamiltonian projected to the lower-band subspace, where we will keep only the processes up to order $t/\Omega$. With these considerations, we can restrict the perturbation to be
% we obtain
\begin{align}
H_\mathrm{int}^{(1)} = \sum_j &\Bigg[
C_0^2S_0^2\frac{U_{\uparrow\uparrow}+U_{\downarrow\downarrow} - 2U_{\uparrow\downarrow}}{2} c_j\dgr c_j\dgr b_j b_j + \Big[ C_0 S_0^3(-U_{\uparrow\uparrow} + U_{\uparrow\downarrow}) + C_0^3 S_0(U_{\downarrow\downarrow} - U_{\uparrow\downarrow}) \Big] c_j\dgr b_j\dgr b_j b_j \cr
+& i\frac{\alpha}2 C_0 S_0(U_{\uparrow\uparrow} + U_{\downarrow\downarrow} - 2 U_{\uparrow\downarrow})\left( c_j\dgr b_{j+1}\dgr b_j b_j - c_j\dgr b_{j-1}\dgr b_j b_j \right) \cr
+& i\frac{\alpha}{2} \left(\left(\frac{C_0^3}{S_0}+\frac{S_0^3}{C_0}-2C_0S_0\right)U_{\uparrow\downarrow}+2C_0S_0(U_{\downarrow\downarrow}+U_{\uparrow\uparrow})\right)\left( - c_j\dgr b_j\dgr b_j b_{j+1} + c_j\dgr b_j\dgr b_j b_{j-1} \right) \cr
-& i\frac{\alpha}2 \left( \frac{S_0^3}{C_0}U_{\uparrow\uparrow} + \frac{C_0^3}{S_0}U_{\downarrow\downarrow} + 2 C_0S_0 U_{\uparrow\downarrow}\right)\left( c_{j+1}\dgr b_{j}\dgr b_j b_j - c_{j-1}\dgr b_{j}\dgr b_j b_j \right) \cr 
-& i\frac{\alpha}2 \Big[S_0^2( U_{\downarrow\downarrow}-U_{\uparrow\downarrow}) -C_0^2( U_{\uparrow\uparrow}-U_{\uparrow\downarrow}) \Big]\left(c_j\dgr c_j\dgr b_j b_{j+1} - c_j\dgr c_j\dgr b_j b_{j-1}\right) 
\Bigg] \cr
+& i\frac{\alpha}2 \Big[ S_0^2 (U_{\uparrow\uparrow}-U_{\uparrow\downarrow}) - C_0^2 (U_{\downarrow\downarrow}-U_{\uparrow\downarrow})\Big]\left(c_{j+1}\dgr c_j\dgr b_j b_j - c_{j-1}\dgr c_j\dgr b_j b_j\right) + \mathrm{H.c.}
\Bigg] + O\left( (t/\Omega)^2 \right), 
\end{align}

For this last step, we set $\delta / \Omega \rightarrow 0$, for which $C_0 = S_0 = 1/\sqrt{2}$. This choice simplifies the resulting expression, since lower- and upper-band dressed states interact locally with the same strength. We can then approximate the energy gap of the excited states to multiples of the single-particle gap
\begin{equation}
\Delta \sim \bra{0}  \left( c_j H_\mathrm{n.i.} c_j^\dagger - b_j H_\mathrm{n.i.} b_j^\dagger \right) \ket{0} = \frac{1}{L}\sum_q (\epsilon_{q,+}-\epsilon_{q,-}),
\end{equation}
which for $\Omega \gg t$ is given by $\Delta \sim \Omega$. With these considerations, we obtain
\begin{align}\label{Seq_int_ham_lower_band_pert}
H_\mathrm{int}^\mathrm{eff} \simeq \sum_j \Big[& \frac{U_\mathrm{eff}^{(0)}+2U_\mathrm{eff}^{(2)}+2W^{(2)} }2 :n_j n_j:  + (\lambda^{(0)}+\lambda^{(2)}) : \bar{j}_j n_j : \cr 
& + W^{(2)} :n_j n_j n_j: + \Lambda^{(2)} : \bar{j}_j n_j n_j: + \Gamma^{(2)} : j_j n_{j+1} n_j:   \Big] ,
\end{align}
with the following expression for the perturbative coefficients
\begin{align}\label{Seq_int_ham_coeffs_pert}
U_\mathrm{eff}^{(2)} &=  -\frac{(U_{\uparrow\uparrow}+U_{\downarrow\downarrow} -2 U_{\uparrow\downarrow})^2}{64\Omega} ,\cr
W^{(2)} &= -\frac{(U_{\downarrow\downarrow}-U_{\uparrow\uparrow})^2}{16\Omega},\cr
\lambda^{(2)} &= -\frac{t\sin(\gamma/2)}{4\Omega^2}(U_{\downarrow\downarrow}-U_{\uparrow\uparrow})\left[U_{\downarrow\downarrow}+U_{\uparrow\uparrow} +\frac{U_{\uparrow\uparrow} + U_{\downarrow\downarrow} - 2U_{\uparrow\downarrow}}{4}\right] \cr
\Lambda^{(2)}  &= -\frac{t\sin(\gamma/2)}{4\Omega^2}(U_{\downarrow\downarrow}-U_{\uparrow\uparrow})\left[U_{\downarrow\downarrow}+U_{\uparrow\uparrow} +\frac{U_{\uparrow\uparrow} + U_{\downarrow\downarrow} - 2U_{\uparrow\downarrow}}{2}\right] \cr
\Gamma^{(2)} &= \frac{t\sin(\gamma/2)}{8\Omega^2}(U_{\downarrow\downarrow}-U_{\uparrow\uparrow})(U_{\uparrow\uparrow} + U_{\downarrow\downarrow} + 2U_{\uparrow\downarrow}).
\end{align}
Finally, by adding the interacting Hamiltonian \eqref{Seq_int_ham_lower_band_pert} to the non-interacting contribution \eqref{Seq_square_ham_ni_truncated_2}, we can write the effective lower-band dressed many-body Hamiltonian as
\begin{align}
H_{-} = \sum_{j}\Big[ &(-t_1 b_{j+1}^\dagger b_{j} - t_2 b_{j+2}^\dagger b_{j} + \mathrm{H.c.}) + \frac{U_\mathrm{eff}}2 :n_j n_j: + t_1\theta  :\bar{j}_j n_j: \cr
& + W :n_j n_j n_j : + \Lambda : \bar{j}_j n_j n_j: + \Gamma : j_j n_{j+1} n_j: \Big],\label{eq_lower_band_full}
\end{align}
where we have relabeled $t_1\theta = \lambda^{(0)}+\lambda^{(2)}$, $W = W^{(2)}$, $\Lambda = \Lambda^{(2)}$ and $\Gamma = \Gamma^{(2)}$. As expected, the coefficients $\theta$, $\Lambda$, $\Gamma$ and $W$ all vanish for spin-symmetric bare collisions, i.e., when $U_{\uparrow\uparrow} = U_{\downarrow\downarrow} = U_{\uparrow\downarrow}$, which was the case investigated in \cite{Barbiero_Cabedo_PRR_2023_frustrated}, where the same flux ladder maps onto a triangular geometry with no chiral interactions. Hamiltonian \eqref{eq_lower_band_full} holds provided the dressed band gap remains the dominant energy scale, $\Omega \gg t$ and $\Omega \gg U_{\sigma\sigma'}$, having retained all contributions up to second order in $1/\Omega$. Under the further assumptions made in Sec.~\ref{sec_chiral}, namely the low-density regime, where the normal-ordered three-body processes proportional to $W$, $\Lambda$ and $\Gamma$ are negligible, and the limit $t/\Omega \to 0$, where $t_2$ vanishes, Hamiltonian \eqref{eq_lower_band_full} reduces to \eqref{eq_lower_band_Ham_low_dens_not2} with the renormalized coupling
\begin{align}
& U_\mathrm{eff} = U_\mathrm{eff}^{(0)}+2U_\mathrm{eff}^{(2)}+2W^{(2)}.\label{Seq_Utheta_2ndpt}
\end{align}
Relevantly, $t_1 \theta$ is the lattice counterpart of the chiral coupling of the chiral BF theory, which the AHM recovers in the continuum limit with a coupling $2\sin(\theta)$ \cite{Bonkhoff_PhysRevLett_2021_linear_anyons}.

\section{Anyonic statistics at arbitrary chiral coupling}\label{Appendix_B_statistics}

In the main text we have shown that, to first order in $\theta$, the Raman anyon-Hubbard model \eqref{eq_lower_band_Ham_theta} reduces to the AHM \eqref{eq_lower_band_AHM}, whose Jordan-Wigner transform describes anyons with the constant statistical phase $\theta$. Here, we show that the correspondence generalizes to arbitrary values of $\theta$, and results in anyonic particles with a density-dependent statistical phase and tunneling with a density-dependent amplitude.

We start by rewriting the tunneling amplitude of Hamiltonian \eqref{eq_lower_band_Ham_theta} as
\begin{equation}\label{Seq_polar_form}
1 - i\theta\bar{n}_j = T_j\, e^{-i A_j}, \qquad
T_j = \sqrt{1 + (\theta\bar{n}_j)^2}, \qquad
A_j = \arctan(\theta\bar{n}_j).
\end{equation}
This writing is an operator identity, because $\bar{n}_j = (n_j + n_{j+1})/2$ is a function of commuting number operators, therefore $\cos A_j = 1/T_j$ and $\sin A_j = \theta\bar{n}_j/T_j$. The kinetic term thus reads $-t_1 b_{j+1}^\dagger T_j e^{-iA_j} b_j + \mathrm{H.c.}$, meaning that a dressed particle picks up the phase $A_j$ when it tunnels across sites with an amplitude $T_j$. For $\theta\bar{n}_j \ll 1$ one recovers $A_j \simeq \theta\bar{n}_j$ and $T_j \simeq 1$, that is, the density-dependent Peierls phase and the tunneling in the JW-transformed AHM \eqref{eq_lower_band_AHM}. Instead, for $\theta\bar{n}_j \gg 1$ the phase saturates, $A_j \to \pi/2$, while the amplitude grows linearly, $T_j \simeq \theta\bar{n}_j$.

Next, we use $f(n_j)\,b_j^\dagger = b_j^\dagger f(n_j+1)$ and the fact that $\bar{n}_j$ contains $n_{j+1}/2$ to obtain
\begin{equation}\label{Seq_reorder_T}
b_{j+1}^\dagger\, T(\bar{n}_j) = T(\bar{n}_j - \tfrac{1}{2})\, b_{j+1}^\dagger .
\end{equation}
and use this result to move the tunneling amplitude to the left of the creation operator in the kinetic Hamiltonian.

We now look for a transformation $a_j = e^{i\Phi_j} b_j$, with $\Phi_j$ a function of the number operators, that removes the phase $A_j$ from the tunneling. Since $a_{j+1}^\dagger a_j = b_{j+1}^\dagger e^{i(\Phi_j - \Phi_{j+1})} b_j$, this amounts to the single requirement
\begin{equation}\label{Seq_string_condition}
\Phi_{j+1} - \Phi_j = A_j ,
\end{equation}
which is solved by
\begin{equation}\label{Seq_string}
\Phi_j = \frac{1}{2}\left(\sum_{k<j} A_k - \sum_{k\geq j} A_k\right).
\end{equation}
Indeed, all sites cancel in the difference $\Phi_{j+1} - \Phi_j$ except the bond $j$, which contributes with $-A_j/2$ from $\Phi_j$ and with $+A_j/2$ from $\Phi_{j+1}$, and with the subtraction reversing the sign of the former. In the limit of small $\theta$, where $A_k \simeq \theta\bar{n}_k$, the two bonds adjacent to site $j$ contribute with opposite signs, the dependence on $n_j$ drops out, and Eq.~\eqref{Seq_string} recovers the JW string that maps the AHM onto bosons.

The redefined operators have the density of the original ones, $a_j^\dagger a_j = b_j^\dagger b_j = n_j$, so that the onsite interactions are left unchanged. Bringing the density-dependent modulus back between the tunneling operators, the Raman anyon-Hubbard model \eqref{eq_lower_band_Ham_theta} expressed in terms of the $a_j$ reads
\begin{equation}\label{Seq_H_anyons}
H_{-} = \sum_j \left[-t_1\left(a_{j+1}^\dagger \sqrt{1+(\theta\bar{n}_j)^2}\, a_j + \mathrm{H.c.}\right) + \frac{U_\mathrm{eff}}{2} n_j (n_j-1)\right].
\end{equation}
The modified anyonic statistics of our model is unveiled by expressing the bosonic commutation relation $[b_i, b_j^\dagger] = \delta_{ij}$ in terms of the new fields. Defining the phase increment produced by the addition of one particle at site $j$ as
\begin{equation}\label{Seq_vartheta}
\vartheta_j = \frac{1}{2}\sum_{l\,=\,j-1}^{j}\Big[\arctan\big(\theta(\bar{n}_l + \tfrac{1}{2})\big) - \arctan\big(\theta\bar{n}_l\big)\Big],
\end{equation}
where the sum runs over the two bonds to which the site belongs, the transformed operators obey the generalized commutation relations
\begin{equation}\label{Seq_anyon_commut_general}
a_i a_j^\dagger - a_j^\dagger\, e^{i\Delta_{ij}}\, a_i = \delta_{ij},
\qquad
\Delta_{ij} = \mathrm{sgn}(i-j)\left(\vartheta_i + \vartheta_j\right),
\end{equation}
and, for neighbouring sites, the bond shared by $i$ and $j$ contributes to both increments. The exchange phase $\Delta_{ij}$ is thus not a number but an operator. The statistical phase is then read off in the Fock sector on which it acts, that is, with site $i$ emptied by $a_i$ and site $j$ occupied. The dressed particles of our Raman anyon-Hubbard model are therefore equivalent to anyons with a density-dependent tunneling and whose statistical phase depends on the local density.

The transformation leaves the tunneling amplitude untouched, because the anyons of \eqref{Seq_H_anyons} tunnel with the same modulus $T_j$ as the Raman-dressed bosons, which grows without bound with the chiral coupling. Instead, the statistical phase is bounded similarly to the tunneling phase $A_j$ of the dressed bosons, and a few limits suffice to characterize the former. For $\theta\bar{n} \ll 1$ all the increments in \eqref{Seq_vartheta} reduce to $\theta/2$, and one recovers the constant angle of the AHM, $\Delta_{ij} \to \mathrm{sgn}(i-j)\,\theta$. For two particles occupying distinct, non-neighbouring sites $i$ and $j$, Eq.~\eqref{Seq_anyon_commut_general} instead gives the exact result
\begin{equation}\label{Seq_two_particle_phase}
\Delta_{ij} = \mathrm{sgn}(i-j)\arctan\theta ,
\end{equation}
independently of their separation and for any value of $\theta$. In this case, the statistical phase grows monotonically with the chiral coupling and saturates at $\pi/2$, so it mirrors the tunneling phase $A_j$. When $\theta(\bar{n}-1/2) \gg 1$ both arctangents in \eqref{Seq_vartheta} are saturated and their difference vanishes as $\Delta_{ij} \simeq \mathrm{sgn}(i-j)/[\theta(\bar{n}^2-1/4)]$, that is, as $1/\theta$ at fixed filling and as $1/\bar{n}^2$ at fixed coupling. In this regime, such a phase contrasts the tunneling phase $A_j$, as the latter grows monotonically towards $\pi/2$, whereas the statistical phase, being built from differences of tunneling phases, is bound to vanish once $A_j$ saturates.

\section{Two-body Hamiltonian}\label{Appendix_C_TwoBodyH}
In this Appendix, we include the explicit expressions of the two-body Hamiltonians studied in Sec.\,\ref{sec_bound_states}. Within the lower-band theory, Hamiltonian \eqref{eq_lower_band_Ham_theta} in the two-particle subspace with center-of-mass momentum $q$ has the matrix representation
\begin{align}
   \hat{H}_{r,r'}(q)= \bra{e_{q,r'}} H_\mathrm{d} \ket{e_{q,r}} = &U_{\mathrm{eff}}\delta_{r,0}\delta_{r',0}
-\left(\delta_{r',r+1}+\delta_{r'+1,r}\right)\left(\beta_1 + \left(\sqrt{2}\rho-\beta_1\right)\left(\delta_{r,0}+\delta_{r',0}\right)\right)\nonumber\\&
-\left(\delta_{r',r+2}+\delta_{r'+2,r}\right)\beta_2\left(1 + \left(\sqrt{2}-1\right)\left(\delta_{r,0}+\delta_{r',0}\right)\right),
\label{eq:H_AHM1D_2particle}
\end{align}
for $\beta_1=2t_1\cos(q/2)$, $\rho=\beta_1(1-\theta\tan(q/2)/2)$ and $\beta_2 = 2t_2\cos(q)$. 

Similarly, the two-body Hamiltonian of the flux ladder model \eqref{eq_ham_flux_ladder} can be written as a block-matrix
\begin{align}
    H=\left[
\begin{array}{c|c|c|c}
\hat{V}_0 & \hat{K}_0 &\hat{0}&\cdots \\ \hline
\hat{K}_0^\dagger & \hat{V} & \hat{K} & \ddots\\
\hat{0} & \hat{K}^\dagger & \ddots &\ddots \\ \vdots & \ddots &\ddots & \ddots
\end{array}\right],
\end{align}
where every submatrix $\hat{V}$ spans the spin space at a fixed relative coordinate $r$. Their explicit expressions are given by
\begin{align}
&\hat{V}_0=\begin{bmatrix}U_{\uparrow\uparrow}-\delta &\frac{\Omega}{\sqrt{2}}&0\\ \frac{\Omega}{\sqrt{2}}&U_{\uparrow\downarrow}&\frac{\Omega}{\sqrt{2}}\\0&\frac{\Omega}{\sqrt{2}}&U_{\downarrow\downarrow}+\delta\end{bmatrix},&\hat{K}_0=-2t\begin{bmatrix}\sqrt{2}\cos{\frac{q+\gamma}{2}} & 0 & 0 & 0\\ 0& e^{i\gamma/2}\cos{\frac{q}{2}} & e^{-i\gamma/2}\cos{\frac{q}{2}} & 0\\ 0& 0 & 0 &\sqrt{2}\cos{\frac{q-\gamma}{2}}\end{bmatrix}\\
&\hat{V}=\begin{bmatrix}-\delta &\frac{\Omega}{2}&\frac{\Omega}{2}&0\\ \frac{\Omega}{2}&0&0&\frac{\Omega}{2}\\ \frac{\Omega}{2}&0&0&\frac{\Omega}{2}\\0&\frac{\Omega}{2}&\frac{\Omega}{2}&\delta\end{bmatrix},&\hat{K}=-2t\begin{bmatrix}\cos{\frac{q+\gamma}{2}} & 0 & 0 & 0\\ 0& e^{i\gamma/2}\cos{\frac{q}{2}} &0&0\\ 0&0 & e^{-i\gamma/2}\cos{\frac{q}{2}} & 0\\ 0& 0 & 0 &\cos{\frac{q-\gamma}{2}}\end{bmatrix}.
\end{align}
The first block is a $3\times 3$ in size, since at $r=0$ the two particles sit at the same lattice sites. Hence, $\ket{e_{q,0}^{\uparrow,\downarrow}}$ is not distinguishable from $\ket{e_{q,0}^{\downarrow,\uparrow}}$.

\bibliography{references}{}
\bibliographystyle{apsrev4-1}

\end{document}